\documentclass[twocolumn]{aastex63}

\accepted{March 25, 2020}
\submitjournal{AJ}

\shorttitle{Horiuchi et al.}
\shortauthors{Horiuchi et al.}

\begin{document}

\title{A Comparison of Properties of Quasars with and without Rapid Broad Absorption Line Variability}

\author[0000-0001-5925-3350]{Takashi Horiuchi}
\affil{Ishigakijima Astronomical Observatory, 
National Astronomical Observatory of Japan, 
National Institutes of Natural Sciences, 
1024-1 Arakawa, Ishigaki, Okinawa, 907-0024, Japan}

\author{Tomoki Morokuma}
\affil{Institute of Astronomy, Graduate School of Science, 
The University of Tokyo, 
2-21-1, Osawa, Mitaka, Tokyo 181-0015, Japan}

\author{Toru Misawa}
\affil{School of General Education, 
Shinshu University, 3-1-1 
Asahi, Matsumoto, Nagano 390-8621, Japan}

\author[0000-0001-8221-6048]{Hidekazu Hanayama}
\affiliation{Ishigakijima Astronomical Observatory, 
National Astronomical Observatory of Japan, 
National Institutes of Natural Sciences, 
1024-1 Arakawa, Ishigaki, Okinawa, 907-0024, Japan}

\author{Toshihiro Kawaguchi}
\affiliation{Department of Economics, 
Management and Information Science, 
Onomichi City University, 
Hisayamada 1600-2, Onomichi, Hiroshima 722-8506, Japan}



\begin{abstract}

We investigate the correlation between rest-frame UV flux variability of broad 
absorption line (BAL) quasars and their variability in BAL equivalent widths (EWs) in a 
various timescale from $<10$~days to a few years in the quasar rest-frame. We 
use the data sets of BAL EWs taken by the Sloan Digital Sky Survey Reverberation 
Mapping (SDSS-RM) project and photometric data taken by the intermediate Palomar 
Transient Factory (iPTF) in $g$ and $R$-bands and the Panoramic Survey Telescope and Rapid 
Response System (Pan-STARRS) in $grizy$ bands. Our results are summarized as below; (1) the 
distributions of flux variability versus BAL variability show weak, moderate, or a strong positive 
correlation, (2) there is no significant difference in flux variability amplitudes between BAL 
quasar with significant short timescale EW variability (called class S1) and without (class S2), 
(3) in all time scales considered in this paper, the class S1 quasars show systematically 
larger BAL variability amplitudes than those of the class S2 quasars, and (4) there are possible correlations 
between BAL variability and physical parameters of the quasars such as black hole masses 
(moderate positive), Eddington ratios, and accretion disk temperature (strong negative) in the 
class S2 quasars. These results indicate that the BAL variability requires changing in the ionizing 
continuum and an ancillary mechanism such as variability in X-ray shielding gas located 
at the innermost region of an accretion disk.   

\end{abstract}



\section{Introduction} \label{sec:intro}
Quasar outflows ejected from their accretion disk around the supermassive 
black holes (SMBHs) are thought to be an important element of a feedback from 
active galactic nuclei since they have following roles; (i) radiatively- and/or magnetically-driven 
winds eject angular momentum from the quasar accretion disk
\citep{1982MNRAS.199..883B,1995ApJ...451..498,2000ApJ...543..686P}, 
(ii) they carry large amounts of energy and metal, then contributing to the chemical evolution of 
the host galaxy \citep{2007A&A...463..513M,2005Natur.433..604D}, and (3) they 
regulate star formation in nearby interstellar and intergalactic regions. 
Thus, outflows are important in that they give insights for the co-evolution mechanism 
between central SMBH and its host galaxy \citep{2000ApJ...539L..9F,2015ApJS...216..4S}.

Quasar outflows are usually detected as quasar absorption lines (QALs) 
in rest-frame UV spectra (intrinsic QALs\footnote{Generally, QALs are classified as 
$intervening$ QALs, which originate in intervening galaxies or the intergalactic medium.} 
hereafter). Intrinsic QALs are classified into following three 
types; Broad absorption lines \citep[BALs; with FWHMs $>$ 2,000 km s$^{-1}$;][]{1991ApJ...373...23W}, 
mini-BALs (with FWHMs of 500 - 2,000 km s$^{-1}$), and narrow absorption lines 
(NALs; with FWHMs $\leq$ 500 km s$^{-1}$). In overall quasar sample, the detection 
rates of BALs, mini-BALs, and NALs are $\sim 20 \%$, $\sim 5 \%$, and $\sim 50 \%$, respectively, 
which probably depend on the viewing angle on our line of sight to the outflow winds 
\citep{1995ApJ...451..498,2001ApJ...549..133G,2012arXiv12013520E,2012AGN...460..47H}.  
On the other hand, the BAL (or mini-BAL) fraction also depends on Eddington ratio and 
black hole mass of quasars; for AGNs with black hole mass of $\sim 10^{8}M_{\odot}$, line-driven 
winds are effectively launched only from the accretion disks with Eddington ratio of higher than 
$\sim$0.01, which yields BALs/mini-BALs on quasar spectra. 
 \citep[see Figures 1 and 4 in][]{2019AA...630..A94G}.

About 90$\%$ of BALs show time variability within 10 years \citep[e.g.,][]{2008ApJ...675..985G,
2011MNRAS.413..908C}. However, physical mechanisms causing BAL variability are still unclear. 
Nowadays, two prevalent scenarios are proposed as follows: (1) 
outflow clouds moving across our line of sight (hereafter, cloud crossing) and (2) changing 
ionization states in the outflow clouds due to variability in the quasar continuum (hereafter, ionization 
state change). \citet{2012ASPC..460..171P} carried out the time-dependent, axisymmetric
simulations of radiation disk winds with the stable disk radiation and confirmed that the very fast 
unsteady flows of the winds dramatically evolve with time. Indeed, the simulated BAL profile of
\citet{2012ASPC..460..171P} became broader with time, which is consistent to
the cloud crossing scenario. 
Previous observational studies support both the cloud crossing scenario 
\citep[e.g.,][]{2008ApJ...675..985G,2013MNRAS.429.1872C,2012ApJ...757..114F,2012MNRAS.421..L107V,
2016MNRAS.455..136V} and the ionization state change scenario \citep[e.g.,][]{1994PASP...106..548B,
2013ApJ...777..168F,2015ApJ...814..50W}, and there is no clear consensus upon which scenario 
is more favorable for the time variations of BAL features; studies of BAL variability with a 
timescale of a few years can not completely decompose the above two scenarios. 

As a rare case, {C\,{\footnotesize IV}\ }BAL of 
SDSS~J141007.74+541203.3 showed an extremely short time scale 
variability within 1.20 days in the quasar rest-frame \citep{2015ApJ...806..111G}. 
Recently, as a first attempt, \citet{2019ApJ...872..21H}, hereafter H19, systematically 
investigated BAL variability in short timescale ($<$10 days) for 27 BAL quasars 
spectroscopically monitored by the Sloan Digital Sky Survey Reverberation 
Mapping (SDSS-RM) project \citep{2015ApJS...216..4S}, and 15 quasars (55$\%$) had 
exhibited significant {C\,{\footnotesize IV}\ }BAL variability within 10 days in the quasar rest frame. 
This result implies that short time scale BAL variability are common phenomena 
in BAL quasars, since these BAL quasars have similar physical 
properties compared with those of the non-BAL quasars. From the perspective of short timescale 
variability, if the sizes of continuum and broad line region (BLR) have an order of $\sim 0.01$ pc or 
larger, crossing velocity of outflow clouds would excess the speed of light. Thus, the cloud 
crossing scenario is disfavored in explaining short time scale variability. 

In the ionization state change scenario, variability of BALs originates in (i) changes of 
radiation intensity from quasars or (ii) changes in shielding gas that locates around the 
roots of outflows and prevents over-ionization of outflows from intense radiation 
\citep{1995ApJ...451..498,2000ApJ...543..686P}. If flux variations and BAL 
variability correlate (or do not correlate) each other, former 
(or latter) explanation can be supported. However, a detailed correlation of UV flux 
and short-timescale BAL variability is still being debated \citep[e.g.,][]{2019MNRAS.468..2379V}. 

In this paper, we discuss the validity of the ionization state change scenario in short 
timescale using frequently observed equivalent width (hereafter, EW) data of BAL quasars 
provided by H19 and light-curves taken by multiple surveys, from which we can study flux 
variability and BAL variability simultaneously. In Section 2, sample selection and catalogs 
used for variability analysis are presented. Results of light curve analyses along with the BAL 
variability and relevant analysis are described in Section 3. In Section 4, we discuss the validity 
of the ionization state change scenario based on the analysis. We summarize our results in 
Section 5. We discuss the correlation between UV flux variability and BAL variability using light 
curves and BAL EWs. Throughout, we adopt a cosmology with $H_{0}=$70 km s$^{-1}$ Mpc$^{-1}$, 
$\Omega_{m}$=0.27 and $\Omega_{\Lambda}$=0.73. 

\section{Sample Selection and Analysis} \label{sec:data}
\subsection{Sample Selection}
In order to examine the correlation between BAL and UV flux variability on short-time 
scale ($<$10-day in the quasar rest-frame), we use EWs of BALs from the H19 sample provided 
by the SDSS-RM project and photometric UV data taken by the intermediate Palomar 
Transient Factory \citep[iPTF;][]{2009PASP...121..134R} and the Panoramic Survey 
Telescope and Rapid Response System \citep[Pan-STARRS, hereafter PS1;][]{2016arXiv161205560C,
2016arXiv161205243F}. Among 27 BAL quasars of the H19 sample, light curves for 25 BAL quasars 
with 34 {C\,{\footnotesize IV}\ }BALs around the SDSS-RM observing epochs are available from the 
two catalogs. Physical parameters of 25 quasars are listed in Table 1 \citep[see also 
H19 and][]{2019ArXiv190403199G}. 

\subsubsection{Data of EW of BALs}

H19 examined {C\,{\footnotesize IV}\ }BAL variability in timescales of 0.21 to 486 days in 
the rest-frame for 27 {C\,{\footnotesize IV}\ }BAL quasars in the SDSS-RM project. 
These BAL quasars are selected with following criteria; (a) {C\,{\footnotesize IV}\ }broad 
absorption feature on quasar spectra satisfies BI $>$ 0 \citep[Balnicity Index;][]{1991ApJ...373...23W}, 
(b) the median signal-to-noise ratios between 1650 and 1750 ${\rm \AA}$ in the quasar rest-frame 
\citep{2009ApJ...696..924G} are 6 pixel$^{-1}$ or higher per pixel (SN$_{1700} \geq 6$), 
corresponding to $i$-band magnitude of $m_i < 20.4$ (see also Figure 2 of H19), (c) BALs are not 
contaminated by residual sky flux or bad pixels. Consequently, H19 selected 27 {C\,{\footnotesize IV}\ }BAL 
quasars in a redshift range of $1.62<z<3.72$. Among 27 quasars, 10 quasars contain two 
{C\,{\footnotesize IV}\ }BALs on their spectra, to which H19 assigned identifiers [A] referred to the 
higher-velocity BAL and [B] to the lower-velocity BAL. For the rest quasars (i.e. 17 out of the 27 quasars) 
with only one {C\,{\footnotesize IV}\ }BAL, H19 applied identifier [A] to the 
{C\,{\footnotesize IV}\ }BAL (see also Table 1). 

Subsequently, H19 defined BAL variability significance $G$
of {C\,{\footnotesize IV}\ }BAL EWs between epoch pairs ($\Delta$EW), by 

\begin{equation}
G = \frac{\chi^2 - (N-1)}{\sqrt{2(N-1)}},
\end{equation}

\begin{equation}
\chi^2 = \sum^{N}_{i=1} \Biggl( \frac{F_2 - F_1}{\sqrt{\sigma_2^2 + \sigma_1^2}} \Biggr)^2_i ~~,
\end{equation}
where the $\chi^2$ indicates the square of the difference of flux between two epochs 
($F_1$ and $F_2$) divided by combined uncertainty, summed over a specified 
region across $N$ spectral pixels (see also equation (3) in H19). The $G$ values  
in the {C\,{\footnotesize IV}\ }BAL regions (or the identified continuum regions) referred to $G_B$ 
(or $G_C$). If $G_B$ is larger than four (as well as $G_C$ is smaller than two, 
i.e., $G_B > 4$ and $G_C < 2$), H19 consider the variability to be a ‘‘significant". 
As a result, they found that 15 out of the 27 quasars (19 of the 37 {C\,{\footnotesize IV}\ }BALs) exhibit 
significant {C\,{\footnotesize IV}\ }BAL variability within 10 rest-frame days. 
We use spectroscopic data (EW of BALs from MJD 56660 to MJD 57933) of 
quasars sampled by H19. Hereafter, we refer to these quasars as class S1 (or S2) that 
show (or do not show) at least one significant {C\,{\footnotesize IV}\ }BAL 
variability within 10 days in the quasar rest-frame. 

\subsubsection{iPTF Photometry}

The iPTF has used the Samuel Oschin 48-inch 
Schmidt Telescope to perform wide-field surveys for a systematic exploration of 
optical transient objects. Most of the iPTF observation times are $g$ and $R$-band 
filters, and the standard exposure time is 60 seconds, which yields 5$\sigma$ limiting 
magnitude of 20.5 ($R$-band) and 21 ($g$-band)\footnote{https://www.ptf.caltech.edu/page/about}. 
The filter transmissions for the $g$ and $R$-band are shown in Figure 4 
of \citet{2009PASP...121..1395}. The data were processed by photometric/astrometric pipeline 
implemented at the Infrared Processing and Analysis Center (IPAC). We use the PTF third data 
release (DR3)  as photometric data by ${\tt MAG\_AUTO}$ \citep[c.f, SExtractor;][]{SExtractor} 
for the 25 quasars (from MJD 54907 to MJD 57043, including the DR1 and DR2). 
Among the 27 quasars, there are no photometric data of the $g$ and $R$-band for 
two quasars (RM 565 and RM 631) in the H19 sample. 

\subsubsection{PS1 Photometry}

The PS1 has performed 3-day cadence observations and photometry in the $grizy$ bands 
using a 1.8 m telescope \citep{2016arXiv161205560C}. The effective wavelength and blue and 
red edges (i.e., band width) of the $grizy$ bands are listed in Table 4 of \citet{2012ApJ...750..99T}. 
To analyze UV flux variability, we use photometric data of the PS1 DR2 from MJD 55215 to MJD 
56864 for 25 quasars. Since photometric data of the PS1 are given in unit 
of Jy, we convert ‘‘psfFlux''\footnote{https://catalogs.mast.stsci.edu/panstarrs/} to 
AB magnitude of which the zero point is 3631 Jy. 

\subsection{Analysis} 

\subsubsection{Correlation between UV flux and {C\,{\footnotesize IV}\ }BAL variability}

In the study of the correlation between UV flux variability and {C\,{\footnotesize IV}\ }BAL 
variability for the individual quasars, we adopt following procedures; (1) we select the data pair 
of the light curves ($m(t_i)$ at i-th observation epoch $t_i$) and the BAL EWs ($EW(t_m^{'})$ at 
m-th observation epoch $t_m^{'}$) in the closest observation epoch with each other,\footnote{
Throughout, we select the photometric data prior to the BAL EW data.} then the time 
lag of the data pair should be within 2.5 days in the quasar rest-frame (i.e., $|t_m^{'} - t_i| < 2.5$ 
days), which is the median time-lag of short timescale BAL variability listed in Table 5 of H19, 
(2) we use the photometric and BAL EW data pairs whose pair numbers are greater 
than 3 per a quasar, (3) we estimate the Pearson's product-moment correlation coefficients 
between UV flux variability and BAL variability $r$ (for all data points that satisfy the 
conditions (1) and (2)) and $r_{\Delta \tau < 5}$ (for the data points whose rest-frame time 
separation between epochs $\Delta \tau$ is smaller than 5 
days, where $\Delta\tau$ is $t_j - t_i$, and $t_j$ represents j-th 
observation epoch) to investigate the interrelation between flux variability ($\Delta m(\Delta\tau) = 
m(t_j) - m(t_i)$) and BAL EW variability ($\Delta EW(\Delta\tau) = EW(t_n^{'}) - EW(t_m^{'})$). 
  
According to the procedures (1) $-$ (3), we examine distributions of $\Delta m$ versus $\Delta EW$ 
and $\Delta m/\bigl<m\bigr>$ versus $\Delta EW/\bigl<EW\bigr>$ (i.e., data point of $\Delta m$ and 
$\Delta EW$ divided by the average magnitude and EW for corresponding observation epochs) per 
rest-frame time separation $\Delta \tau$ for both axis. In this paper, the criteria of correlation 
coefficients with weak, moderate, and strong correlations correspond to $0.3 < |r| < 0.4$, 
$0.4 < |r| < 0.6$, and $|r| > 0.6$, respectively \footnote{The criterion of 
correlation coefficients is based on \citet{2019MNRAS.468..2379V}.}. If absolute values of 
correlation coefficients are less than 0.3, or the corresponding probabilities (i.e., $p$-values) 
are larger than 0.05, we regard those as no possible correlation. We do not consider photometric 
and EW errors for correlation coefficients. 

\subsubsection{Structure Function Analysis}
The structure function analysis is one of the most popular ways to examine time 
dependence of flux variability amplitude. We analyze the structure function 
for the classes S1 and S2 quasars using the iPTF and the PS1 light curves with 
the definition by \citet{1996ApJ...463..466D},    

\begin{equation}
SF_m(\Delta \tau) = \sqrt{\frac{\pi}{2} \bigl< |\Delta m(\Delta \tau)| \bigr>^2 - \bigl<\sigma_{ij}^2\bigr>},
\end{equation}
where $\bigl<|\Delta m(\Delta \tau)|\bigr>$ and $\bigl<\sigma_{ij}^2\bigr>~(=~\bigl<\sigma_{i}^2
+\sigma_{j}^2\bigr>$) are the average values (denoted in bracket) of flux (magnitude) variability 
amplitude and photometric errors separated by time delay between two observation epochs 
$\Delta \tau$ in the quasar rest-frame. The structure function analysis has advantages of being less 
susceptible to outliers of flux variability amplitude and easily applicable to unevenly sampled 
time series. We also adopt a structure function formula defined by \citet{2007A&A...470..49T} 
for BAL EWs of the class S1 and S2 quasars, 

\begin{eqnarray}
SF_{\rm BAL}(\Delta \tau) = \frac{1}{M} \sum^{M}_{i,j}\Biggl[ UDSF \Biggr], \\
 UDSF  = \sqrt{\frac{\pi}{2}} | {\rm log}~EW(t_j) - {\rm log}~EW(t_i) |, 
\end{eqnarray}
where $EW (t_i)$ and $EW (t_j)$ are EWs of {C\,{\footnotesize IV}\ }BAL on 
observation epochs $t_i$ and $t_j$, and the sum means add up for all $M$ pairs of 
observation epochs. An unbinned discrete structure function 
\citep[e.g.,][]{2007A&A...470..49T,2013A&A...557A..91T} is shown as $UDSF$. 
We estimate $UDSF$ and $SF_{\rm BAL}$ for the 34 individual {C\,{\footnotesize IV}\ } BALs.

\begin{figure*}
\figurenum{1}
\includegraphics[clip, width = 17.7cm, height = 21.5cm]{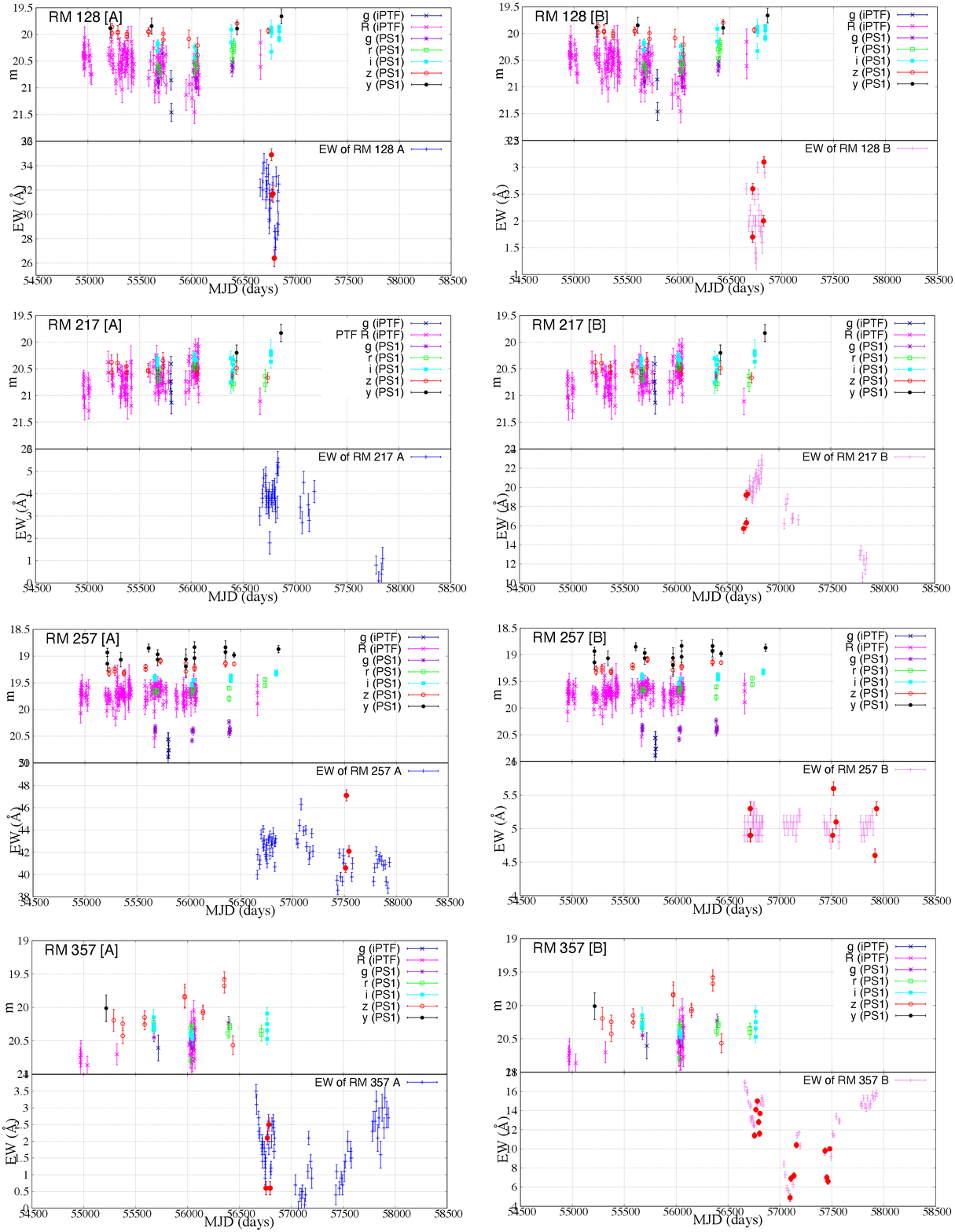}
\caption{Light curves of the iPTF $g$ (navy), $R$ (magenta)-band, and the 
PS1 $g$ (purple), $r$ (green), $i$ (cyan), $z$ (red), and $y$ (black)-band ($top$), 
and EW variability of {C\,{\footnotesize IV}\ }BALs after dividing into A and 
B ($bottom$) for the class S1 quasars which show the significant BAL EW variability 
defined by H19. Red circles indicate data pairs that satisfy criteria of H19 for the 
significant short timescale BAL variability.  \label{fig:f1}}
\end{figure*}

\begin{figure*}
\figurenum{1}
\includegraphics[clip, width = 17.7cm, height = 22.5cm]{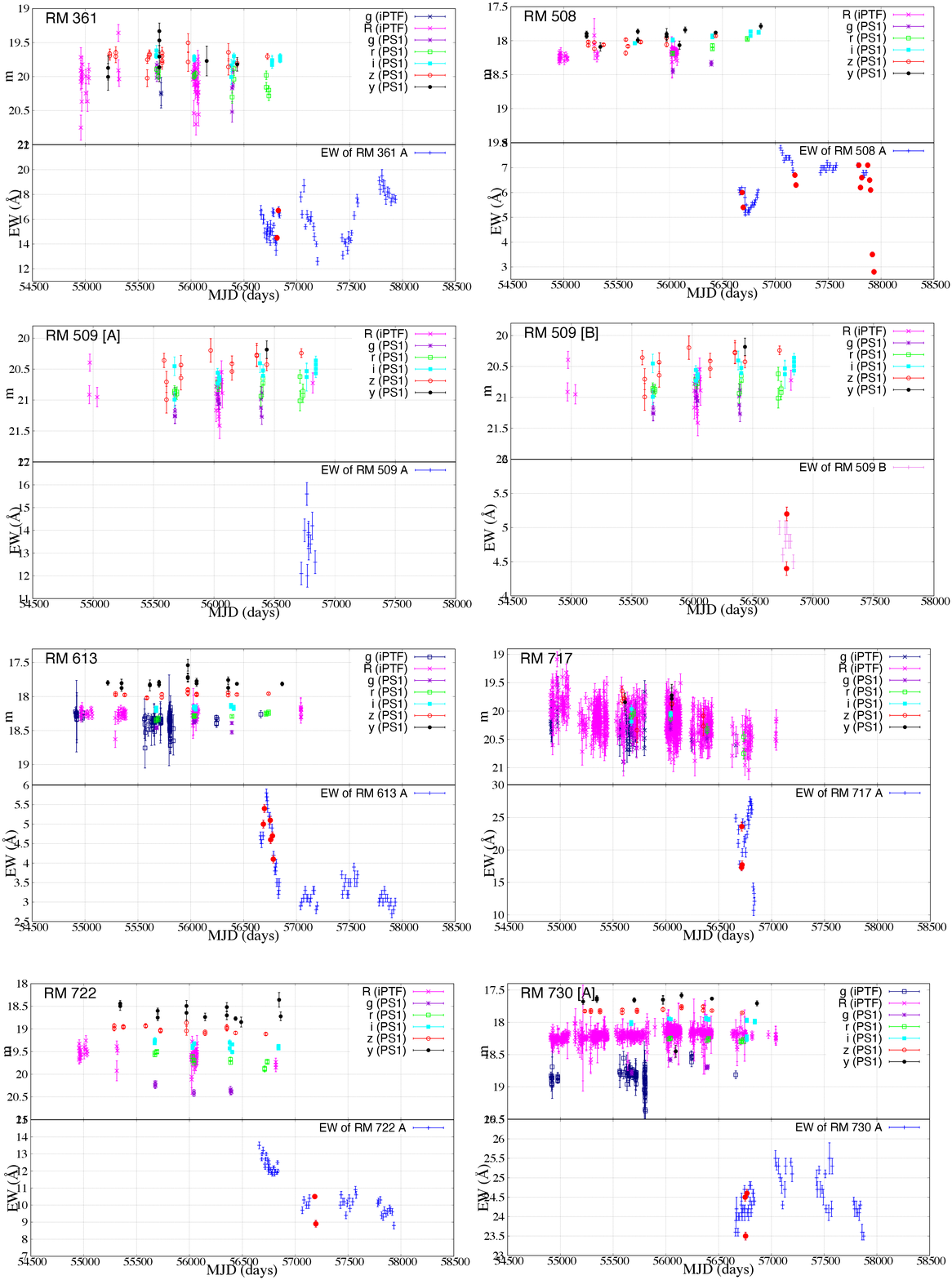}
\caption{$Continued$.}
\end{figure*}

\begin{figure*}
\figurenum{1}
\includegraphics[clip, width = 17.7cm, height = 17.5cm]{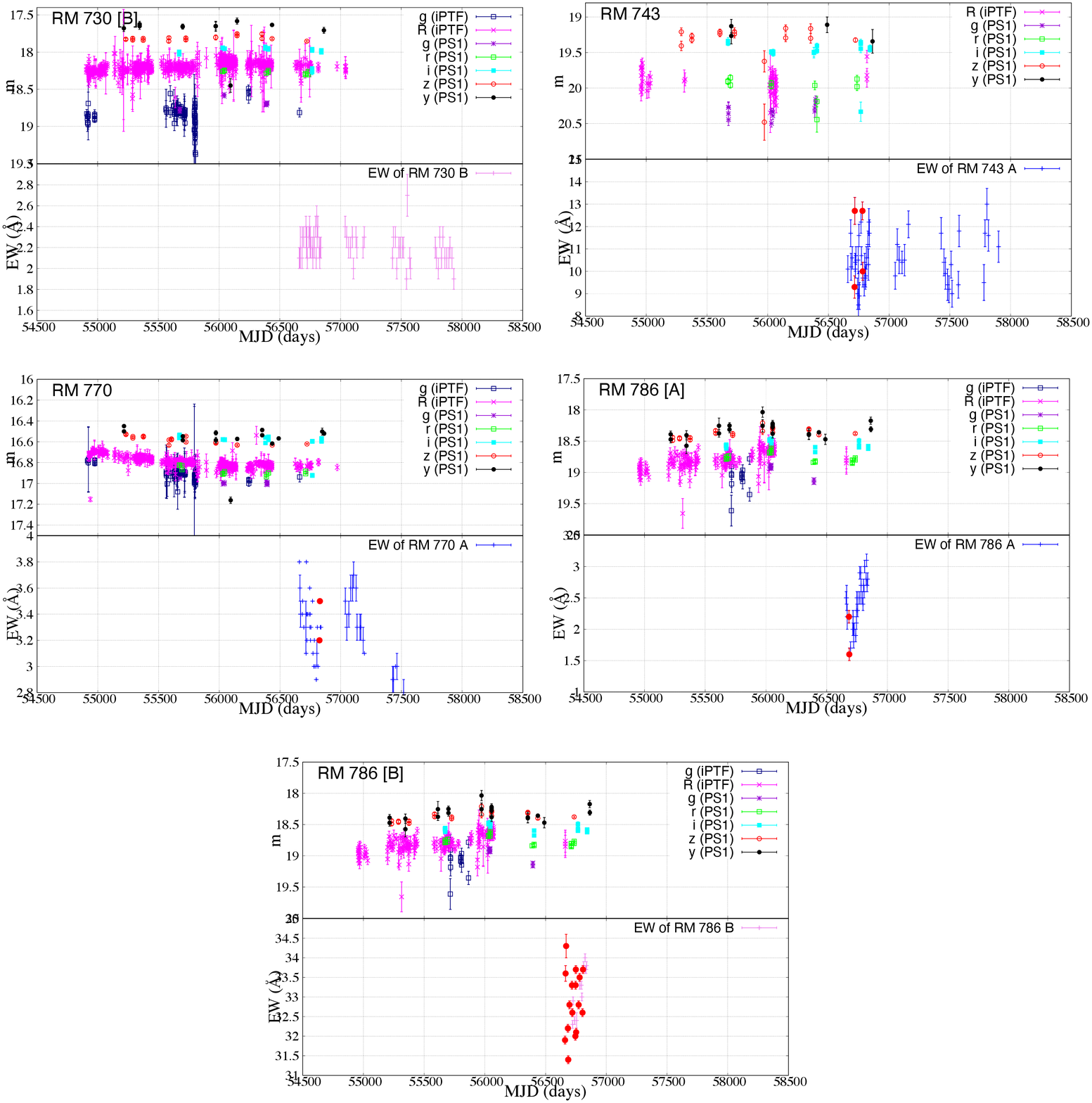}
\caption{$Continued$.}
\end{figure*}

\begin{figure*}
\figurenum{2}
\includegraphics[clip, width = 17.7cm, height = 22.5cm]{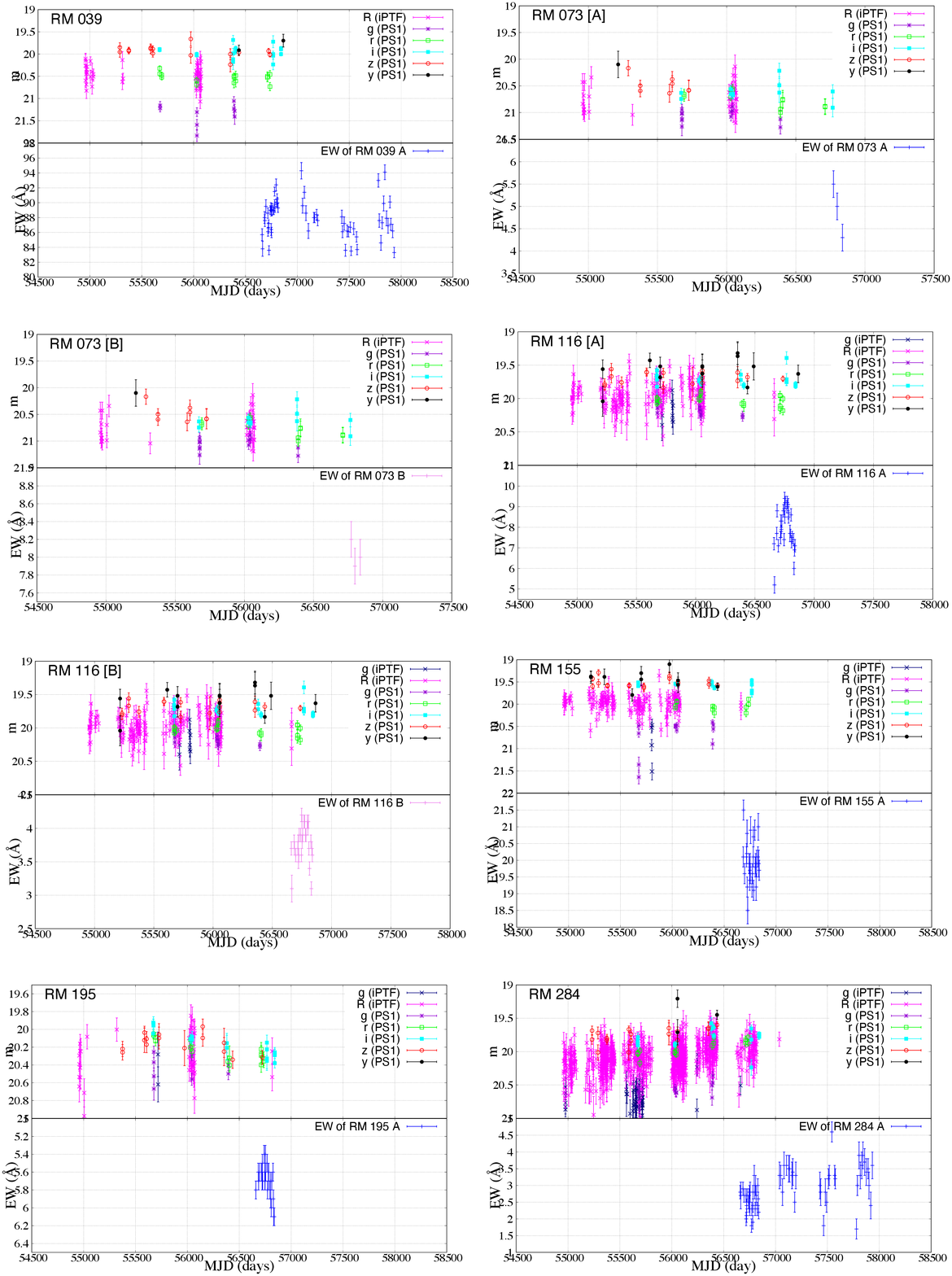}
\caption{The same as Figure 1, but for the class S2 quasars. \label{fig:f2}}
\end{figure*}

\begin{figure*}
\figurenum{2}
\includegraphics[width = 17.7cm, height = 17.5cm]{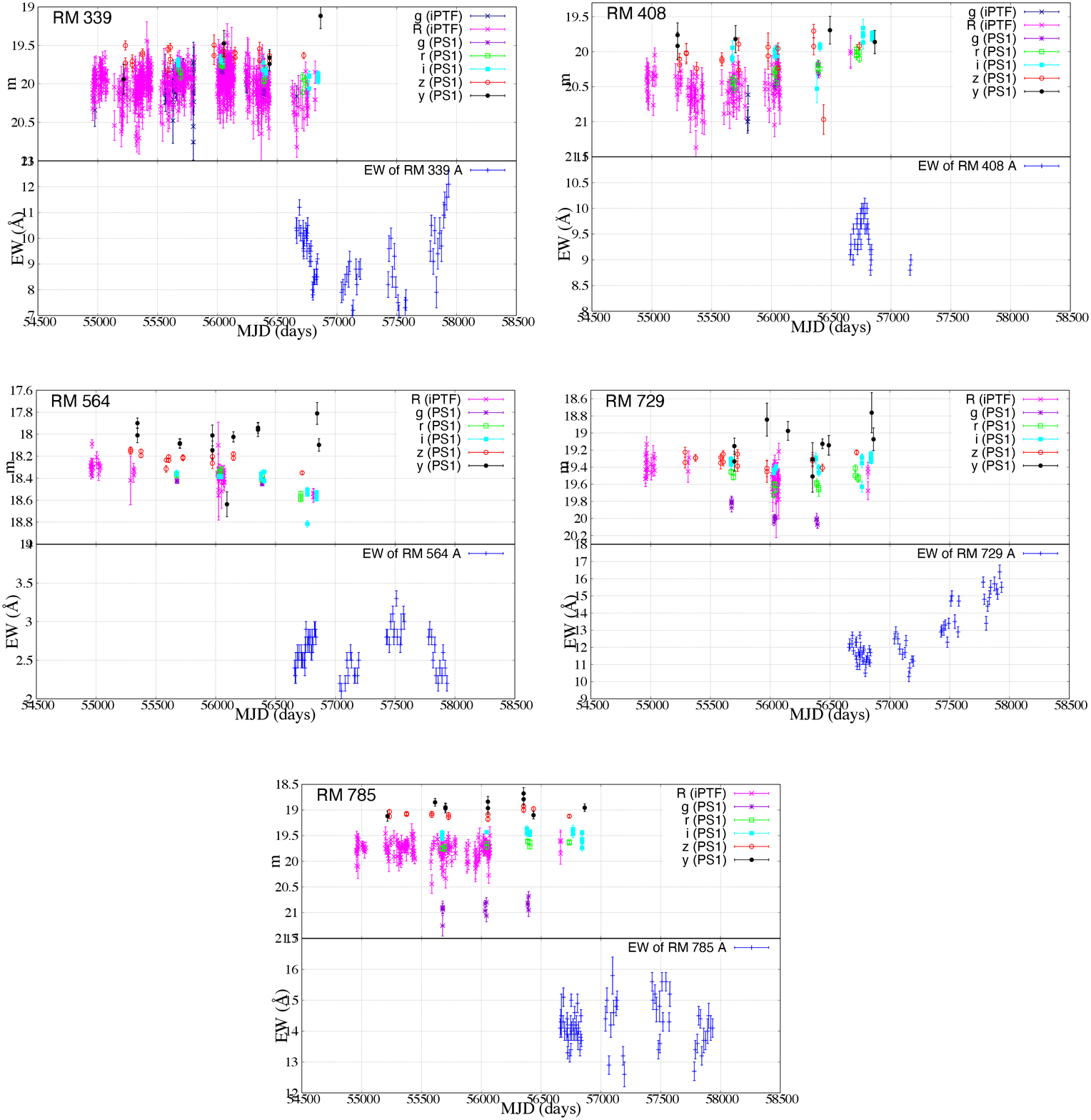}
\caption{$Continued$.}
\end{figure*}

\begin{deluxetable*}{ccCrlccccc}[b!]
\tablecaption{Physical properties of sample quasars \label{tab:mathmode}}
\tablecolumns{8}
\tablenum{1}
\tablewidth{0pt}
\tablehead{
\colhead{SDSS ID} &
\colhead{RM ID} &
\colhead{Redshift} & \colhead{$m_i$\tablenotemark{a}} & \colhead{$N_{\rm obs}$\tablenotemark{b}} &
 \colhead{Identifier\tablenotemark{c}} &
\colhead{Class\tablenotemark{d}} &
\colhead{log $\lambda L_{\lambda 1350}$} \tablenotemark{e} & \colhead{log {$M_{\rm BH}$}}\tablenotemark{f} &
$L_{\rm bol}/L_{\rm E}$\tablenotemark{g}\\
& & & & & & &  \colhead{(erg s$^{-1}$)} &  \colhead{($M_{\odot}$)} 
}
\startdata
J141607.12+531904.8 & RM 039 & 3.08 & 19.77 & 58 & A & S2 & 45.619$\pm$0.003 & 8.48$\pm$0.07 & 0.109$\pm$0.017\\
J141741.72+530519.0 & RM 073 & 3.43 & 20.37 & 3 & A, B & S2 & $\ldots$ & $\ldots$ & $\ldots$ \\
J141432.46+523154.5 & RM 116 & 1.88 & 19.68 & 64 & A, B & S2 & 45.652$\pm$0.001 & 8.90$\pm$0.03 & 0.044$\pm$0.003\\
J141103.17+531551.3 & RM 128 & 1.86 & 20.01 & 50 & A$^*$, B$^*$ & S1 & 45.359$\pm$0.002 & 8.68$\pm$0.05& 0.037$\pm$0.004\\
J141123.68+532845.7 & RM 155 & 1.65 & 19.65 & 41 & A & S2 & 45.364$\pm$0.001 & $\ldots$ & $\ldots$ \\
J141935.58+525710.7 & RM 195 & 3.22 & 20.33 & 59 & A &S2 & $\ldots$ & $\ldots$ & $\ldots$ \\
J141000.68+532156.1 & RM 217 & 1.81 & 20.39 & 41 & A, B$^*$ & S1 & 45.382$\pm$0.002 & 8.67$\pm$0.02 &0.040$\pm$0.001\\
J140931.90+532302.2 & RM 257 & 2.43 & 19.54 & 65 & A$^*$, B$^*$ & S1 & 45.782$\pm$0.005 & 9.19$\pm$0.04 & 0.031$\pm$0.002\\
J141927.35+533727.7 & RM 284 & 2.36 & 20.22 & 63 & A & S2 & 45.642$\pm$0.001 & 9.05$\pm$0.05 & 0.031$\pm$0.003\\
J142014.84+533609.0 & RM 339 & 2.01 & 20.00 & 64 & A & S2 & 45.743$\pm$0.001 & 8.94$\pm$0.01 & 0.050$\pm$0.001\\
J141955.27+522741.1 & RM 357 & 2.14 & 20.23 & 65  & A$^*$, B$^*$ & S1 & $\ldots$ & $\ldots$ & $\ldots$\\
J142100.22+524342.3 & RM 361 & 1.62 & 19.46 & 66 & A$^*$ & S1 & 45.576$\pm$0.001 & $\ldots$ & $\ldots$ \\
J141409.85+520137.2 & RM 408 & 1.74 & 19.63 & 65 & A & S2 & 45.708$\pm$0.001 & 8.47$\pm$0.09&0.137$\pm$0.028\\
J142129.40+522752.0 & RM 508 & 3.21 & 18.12 & 69 & A$^*$ & S1 & 46.919$\pm$1.000 & 10.35$\pm$1.00 & 0.029$\pm$0.095\\
J142233.74+525219.8 & RM 509 & 2.65 & 20.30 & 10  & A, B$^*$  & S1 & $\ldots$ & $\ldots$ & $\ldots$\\
J142306.05+531529.0 & RM 564 & 2.46 & 18.24 & 67 & A & S2 & 46.484$\pm$0.000 & 9.42$\pm$0.01 & 0.091$\pm$0.002\\
J141007.73+541203.4 & RM 613 & 2.35 & 18.12 & 68 & A$^*$ & S1 & 46.591$\pm$0.001 & 9.10$\pm$0.01 & 0.245$\pm$0.005\\
J141648.26+542900.9 & RM 717 & 2.17 & 19.69 & 49 & A$^*$  & S1 & $\ldots$ & $\ldots$ & $\ldots$ \\
J142419.18+531750.6 & RM 722 & 2.54 & 19.49 & 59 & A$^*$ & S1 & 45.799$\pm$0.002 & 9.20$\pm$0.07 & 0.031$\pm$0.005\\
J142404.67+532949.3 & RM 729 & 2.76 & 19.56 & 63 & A & S2 & 46.074$\pm$0.001 & 9.10$\pm$0.01 & 0.074$\pm$0.001\\
J142225.03+535901.7 & RM 730 & 2.69 & 17.98 & 65 & A$^*$, B & S1 & $\ldots$ & $\ldots$ & $\ldots$\\
J142405.10+533206.3 & RM 743 & 1.74 & 19.18 & 49 & A$^*$ & S1 & 45.389$\pm$0.002 & 8.53$\pm$0.01 & 0.057$\pm$0.001\\
J142106.86+533745.2 & RM 770 & 1.86 & 16.46 & 69 & A$^*$ & S1 & 46.948$\pm$0.003 & 9.31$\pm$0.10 & 0.344$\pm$0.007\\
J141322.43+523249.7 & RM 785 & 3.72 & 19.19 & 67 & A & S2 & $\ldots$ & $\ldots$ & $\ldots$\\
J141421.53+522940.1 & RM 786 & 2.04 & 18.63 & 68 & A$^*$, B$^*$ & S1 & $\ldots$ & $\ldots$ & $\ldots$\\
\enddata
\tablenotetext{a}{$i$-band magnitudes obtained from the SDSS DR10.}
\tablenotetext{b}{Number of epochs for spectroscopic observations.}
\tablenotetext{c}{The identifier for higher-velocity (A) and lower-velocity (B) {C\,{\footnotesize IV}\ }BALs. 
‘‘A, B" indicates that the quasar spectrum contains two {C\,{\footnotesize IV}\ }BALs. 
Asterisks indicate the {C\,{\footnotesize IV}\ }BALs with short timescale BAL variability within 
10 days in the quasar rest-frame.}
\tablenotetext{d}{Subsample classes defined in Section 2.1.}
\tablenotetext{e}{Bolometric luminosities estimated by flux at 1350 ${\rm \AA}$. See also \citet{2019ArXiv190403199G}.}
\tablenotetext{f}{Black hole masses in the unit of solar mass. See also \citet{2019ArXiv190403199G}.}
\tablenotetext{g}{Eddington ratios.}
\end{deluxetable*}

\section{Results}

\subsection{Correlation between UV flux and {C\,{\footnotesize IV}\ }BAL variability}
As a first step to examine UV flux variability and BAL variability simultaneously, in Figures 1 
and 2, we show light curves of the iPTF $g, R$ data and the PS1 $grizy$ data with 
BAL EWs ([A] and [B]) for the 34 BALs in the 25 quasars (i.e., excluding 
the data of RM 565 and RM 631) on the same time axis for the classes S1 and S2 quasars. 
In Figures 1 and 2, only 5 quasar/BAL pairs with the iPTF $R$-band 
photometric data satisfy the criteria in Section 2.2.1 (RM 284 [A], RM 339 [A], 
RM 717 [A], RM 730 [A], RM 730 [B], and RM770 [A]; hereafter, $five~quasars$). As a next 
step, we show the relation of the iPTF $R$-band variability $\Delta R$ versus the EW variability 
$\Delta EW$ (Figure 3). In Figure 4, we also indicate $\Delta R/\bigl<R\bigr>$ versus 
$\Delta EW/\bigl<EW\bigr>$ divided by $\Delta \tau$ for the $five~quasars$. 
We denote the details of these distributions for the $five~quasars$ 
in the following sections, including their spectral features (see also APPENDIX A in H19).  
The redshift, i-band magnitudes, and the class (S1 or S2) of each quasar and 
BAL are summarized in the parenthesis. 

\subsubsection{RM 284 [A] ($z=2.36, m_i=20.22, class: S2$)} 
In the RM284 spectrum, an extremely high velocity {C\,{\footnotesize IV}\ }trough is 
detected, and its velocity reaches from $\sim$ 35,000 to $\sim$ 59,000 km s$^{-1}$, 
that makes detection of any {Si\,{\footnotesize IV}\ }absorption in BAL[A] 
unclear. {Al\,{\footnotesize III}\ }absorption also does not clearly appear on the spectrum.

In Figure 3a, the number of data pair $N$ for $R$-band light curve data points 
and {C\,{\footnotesize IV}\ }BAL EWs is 17 (136 combinations of the observation epochs), 
which is the largest numbers in the $five~quasars$. We cannot find correlation 
of the distribution ($r=-0.03$, $r_{\Delta \tau < 5}=0.04$). In Figure 4a, the distribution 
shows a weak or a moderate correlation ($r=0.39$, $r_{\Delta \tau < 5}=0.42$ with 
$p$-values$<$0.04). 
 
\subsubsection{RM 339 [A] ($z=2.01, m_i=20.00, class: S2$)}
This quasar's spectrum contains {Si\,{\footnotesize IV}\ }and {Al\,{\footnotesize III}\ }BALs, 
whose velocity shifts are similar to the {C\,{\footnotesize IV}\ }trough. Figures 3b and 4b 
exhibit no possible correlations of the distributions in any timescales (i.e., $|r|<0.3$ or 
$p$-values$>$0.05). 

\subsubsection{RM 717 [A] ($z=2.17, m_i=19.69, class: S1$)}
The {C\,{\footnotesize IV}\ }BAL of this quasar (BAL[A]) is composed of multiple 
components, and the accompanying {Si\,{\footnotesize IV}\ }BAL also exhibits 
a similar structure. BAL[A] showed the largest EW variability amplitude in the $five~quasars$. 
In Figure 3c, the $\Delta R$ - $\Delta EW$ distribution is shown no correlations 
($r=-0.01$, $r_{\Delta \tau < 5}=-0.07$). In Figure 4c, the correlation coefficient for all of 
data points indicates a weak correlation ($r=0.34$ with a $p$-value$<$0.01). 

\subsubsection{RM 730 [A], [B] ($z=2.69, m_i=17.98, class: S1$)}
The spectrum of RM 730 contains two {C\,{\footnotesize IV}\ }BALs (BAL[A] and BAL[B]) 
separated by $\sim$ 500 km s$^{-1}$ from a red edge of the BAL[A] to a blue edge of the BAL[B], 
Only the BAL[A] exhibits a short timescale (rapid) variability based on criteria as defined in H19. A 
{Si\,{\footnotesize IV}\ }BAL and shallow {Al\,{\footnotesize III}\ }features (not BAL) are detected 
at velocities corresponding to BAL[A]. There are no BAL features of {Si\,{\footnotesize IV}\ }, 
{Al\,{\footnotesize III}\, }, or {P\,{\footnotesize V}\ }at velocities associated to BAL[B]. 

The quasar does not show significant flux variability (i.e., $\Delta R < 3\sigma$ photometric error). 
Figure 3d (for RM 730 [A]) presents EW variability ($\Delta EW$) that are comparable to these 
of RM 284 [A] and RM 339 [A], and Figure 3e (for RM 730 [B]) shows smaller EW variability 
than those of RM 284 [A], RM 339 [A], RM 717 [A], and RM 730 [A]. In Figures 3d, 3e, and 4d,  
there are no possible correlations in the $\Delta R$ --- $\Delta EW$ distributions ($r=0.006$, 
$r_{\Delta \tau < 5}=0.17$ for Figure 3d, $r=0.18$, $r_{\Delta \tau < 5}=0.31$ with 
$p$-value=0.25 for Figure 3e, and $r=-0.06$, $r_{\Delta \tau < 5}=-0.11$ for Figure 4d). 
Meanwhile, Figure 4e exhibits a strong correlation ($r=0.88$, $r_{\Delta \tau < 5}=0.90$ 
with $p$-values$<$0.0001 for RM 730 [B]).

\subsubsection{RM 770 [A]  ($z=1.86, m_i=16.46, class: S1$)}
This quasar is the apparently brightest in the H19 sample and does not contain 
both {Si\,{\footnotesize IV}\ }and {Al\,{\footnotesize III}\ }absorptions. 
The quasar also shows no significant flux variability within $R$-band photometric errors. 
In Figures 3f and 4f, we find no possible correlations of the distributions with $|r|<0.3$ or 
$p$-values$>$0.11. \\

In Figure 4, three of six distributions (i.e., except for RM 339 [A], RM 730 [A], and RM 770 [A]) 
present weak, moderate, or the strong correlation, and we also find that the correlation 
coefficients $r$ and $r_{\Delta \tau < 5}$ are comparable for each BAL of the 
$five~quasars$.

\begin{figure*}
\figurenum{3}
\plotone{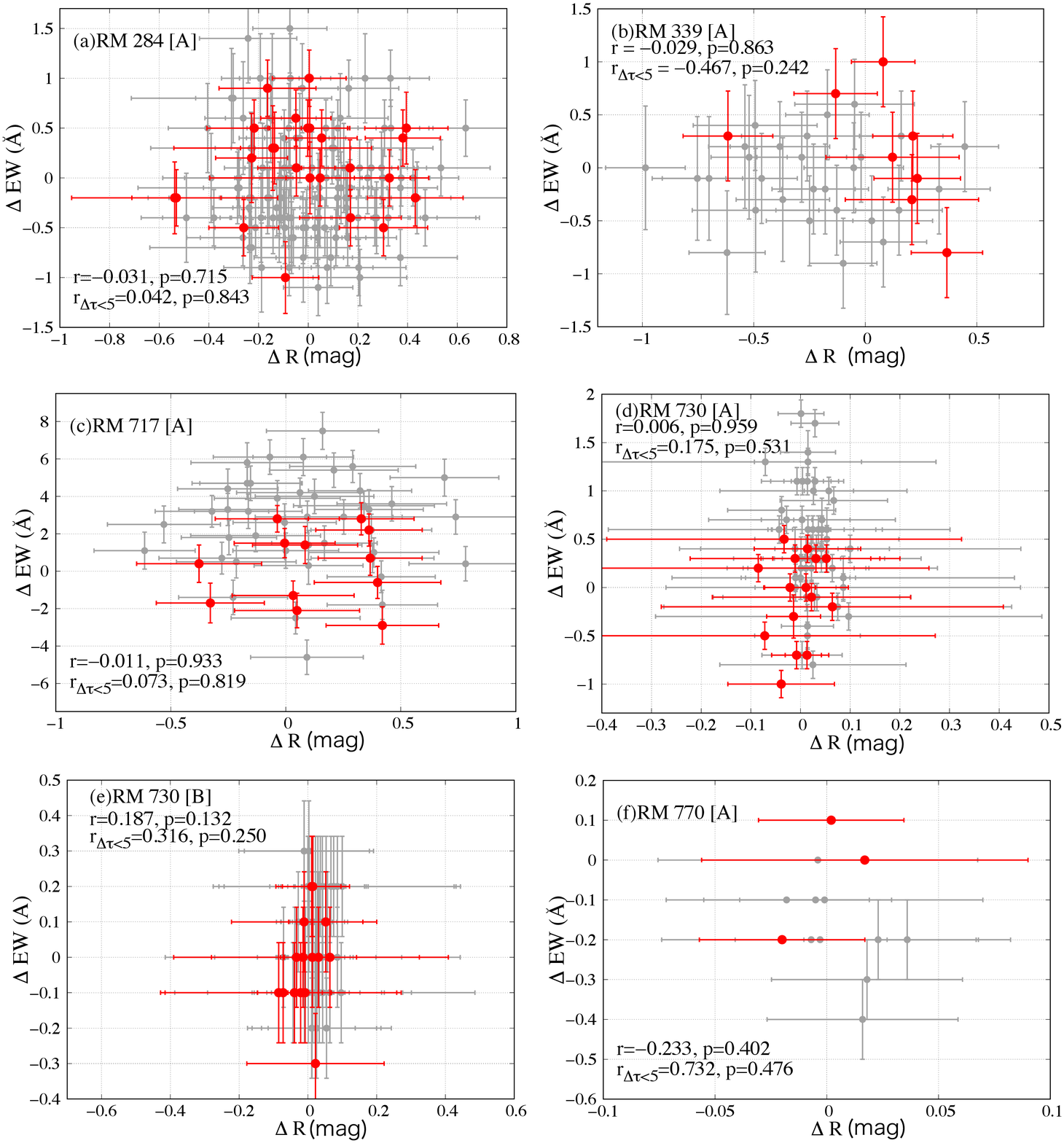}
\caption{Distributions of $\Delta R$ versus $\Delta EW$ for (a) RM 284 [A], 
(b) RM 339 [A], (c) RM 717 [A], (d) RM 730 [A], (e) RM 730 [B], and (f) RM 770 [A]. 
Filled gray and red circles indicate the data points whose time separation 
is larger and smaller than 5 days. The correlation coefficients $r$ (for all of data points) 
and $r_{\Delta \tau < 5}$ (for data point of $\Delta \tau <$ 5 days) with the associated 
$p$-values are described in each panel. 
\label{fig:f3}}
\end{figure*}

\begin{figure*}
\figurenum{4}
 \plotone{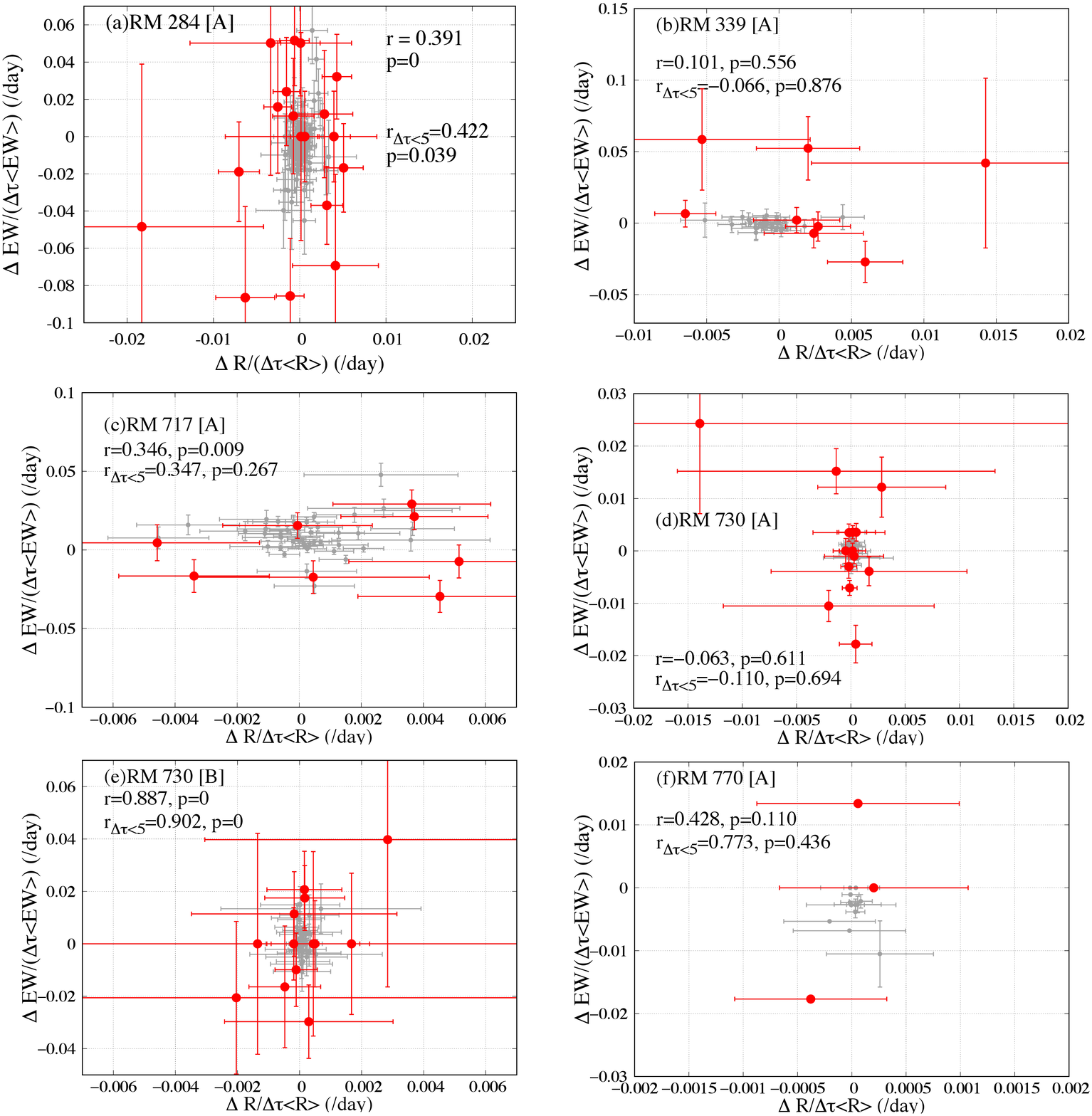}
\caption{The same as Figure 3, but for distributions of $\Delta R/\bigl<R\bigr>$ versus 
$\Delta EW/\bigl<EW\bigr>$ both of which are divided by the rest-frame time 
separations $\Delta \tau$. The $p$-values with zero values indicate $p$-value$<$0.0001.
\label{fig:f4}}
\end{figure*}
 
\subsection{Structure Function Analysis}
The PS1 DR2 data enable detailed structure function analysis 
to compare between the classes S1 and S2 quasars. On the other hand, it is difficult to investigate 
structure function analysis based on eqn. (3) using the iPTF $g, R$-band data due to large 
photometric error, since the average values of photometric error $\bigl<\sigma_{ij}^2\bigr>$ excess 
the mean UV flux variability amplitudes $\bigl< |\Delta m(\Delta \tau)| \bigr>^2$ especially in short 
timescale. 

The rest-frame wavelength range through photometric bands depends on the redshifts of individual 
quasars. Therefore, we calculate structure functions for four bins of the rest-frame wavelength 
with the red-edges of the PS1 $grizy$ bands at the following rest-frame wavelength ($\lambda_{\rm red}$); 
(a) $\lambda_{\rm red} < 2000~{\rm \AA}$, (b) $2000~{\rm \AA} < \lambda_{\rm red} < 2500~{\rm \AA}$, 
(c) $2500~{\rm \AA} < \lambda_{\rm red} < 3000~{\rm \AA}$, and (d) $\lambda_{\rm red} > 3000~{\rm \AA}$, 
corresponding to full wavelength ranges ($\lambda_{\rm blue} <  \lambda < \lambda_{\rm red}$; 
$\lambda_{\rm blue}$ corresponds to blue-edges of the following full wavelength ranges) of (a) $877~{\rm \AA} 
< \lambda < 1991~{\rm \AA}$, (b) $1510~{\rm \AA} < \lambda < 2498~{\rm \AA}$, (c) $2007~{\rm \AA} < 
\lambda < 2989~{\rm \AA}$, and (d) $2603~{\rm \AA} < \lambda < 3820~{\rm \AA}$ for the all of 
photometric data of the 25 quasars, respectively. 

We estimate the structure function for the classes S1 and S2 quasars, including the short 
timescale variability $SF(\Delta\tau<10$ day). In the characterization of structure functions, 
power-law 
\citep[e.g.,][]{1994MNRAS.268..305H,2002ApJS..141...45E,2004ApJ...601..692V,2012ApJ...753..106M},

\begin{equation}
SF(\Delta \tau) = \Biggl(\frac{\Delta \tau}{\Delta \tau_{p}} \Biggr)^\gamma \equiv b  \Delta \tau^\gamma ,
\end{equation}
or a damped random walk \citep[DRW: e.g.,][]{2009ApJ...698..895K,2010ApJ...721..1024M,
2012ApJ...753..106M,2016ApJ...826...118K,2017MNRAS.470..4112G}, 

\begin{equation}
SF(\Delta \tau) =  SF_{\infty} \sqrt{1-{\rm exp}(-\Delta \tau/\Delta \tau_{\rm DRW})}, 
\end{equation}
is usually employed, where $\gamma$ and $\Delta \tau_{p}$ in eqn. (6) are the power-law index and 
the time scale such that $SF(\Delta \tau_{p})$ equals 1 mag. $SF_{\infty}$ and 
$\Delta \tau_{\rm DRW}$ in eqn. (7) are the asymptotic value at $\Delta \tau$ = $\infty$ 
and characteristic timescale, respectively. Figure 5 shows the structure functions of the light curves 
with fitted lines and describing fitted parameters of eqns. (6) and (7). Due to the large uncertainty in 
fitted parameters, we cannot evaluate accurate values for $\Delta \tau_{p}$ and $\Delta \tau_{\rm DRW}$;  
fitting errors of these two parameters exceed their own values. 

Besides the analysis above for the continuum variability, we also apply the same approach to the 
structure function analysis for {C\,{\footnotesize IV}\ }EW variability based on eqns. (4) and (5) 
to these fitting models (Figure 6). In Figure 6b, we introduce another category to the classes S1 
and S2 quasars; {C\,{\footnotesize IV}\ }BALs with significant short timescale variability defined by 
H19 are marked in asterisks in Table 1 ($*$) or not ($None$). In this category, we cannot make a 
quantitative physical parameter $\Delta \tau_p$ with the large uncertainty. As a similar BAL variability 
trend to the classes S1 and S2 quasars, the category $*$ shows the systematically larger BAL 
variability than that of the category “None". The overall trends is that there is no clear difference in UV 
flux variability amplitudes between S1 and S2 (Figure 5), while BAL variability of the class S1 
quasars (or “$*$'') are systematically larger than that of the class S2 (or “None") 
from short to long timescales (Figure 6).  

The flux we monitored in this study has larger wavelength than those of ionizing EUV flux. 
Therefore, we test flux variability dependence on the rest-frame wavelength to clarify a relation 
between the UV and ionizing continuum to {C\,{\footnotesize IV}\ } ions ($h\nu>~$47 eV, 
$\lambda<260{\rm \AA}$). In general, the shorter the wavelength is, the larger quasar variability 
becomes \citep[i.e., bluer-when-brighter (BWB) 
trend; e.g.,][]{1999MNRAS.306..637G,2004ApJ...601..692V,2014ApJ...783..46K}. Then we characterize 
the trend for the class S2 quasars by

\begin{equation}
SF(\lambda) =  a_0\exp{(-\lambda/\lambda_0)}, 
\end{equation}
where $a_0$ and $\lambda_0$ are fitting parameters \citep{2004ApJ...601..692V}. In Figure 7, 
the class S2 quasars show a weak BWB trend in contrast to the class S1 quasars with large 
photometric errors.

\begin{figure*}
\figurenum{5}
 \plotone{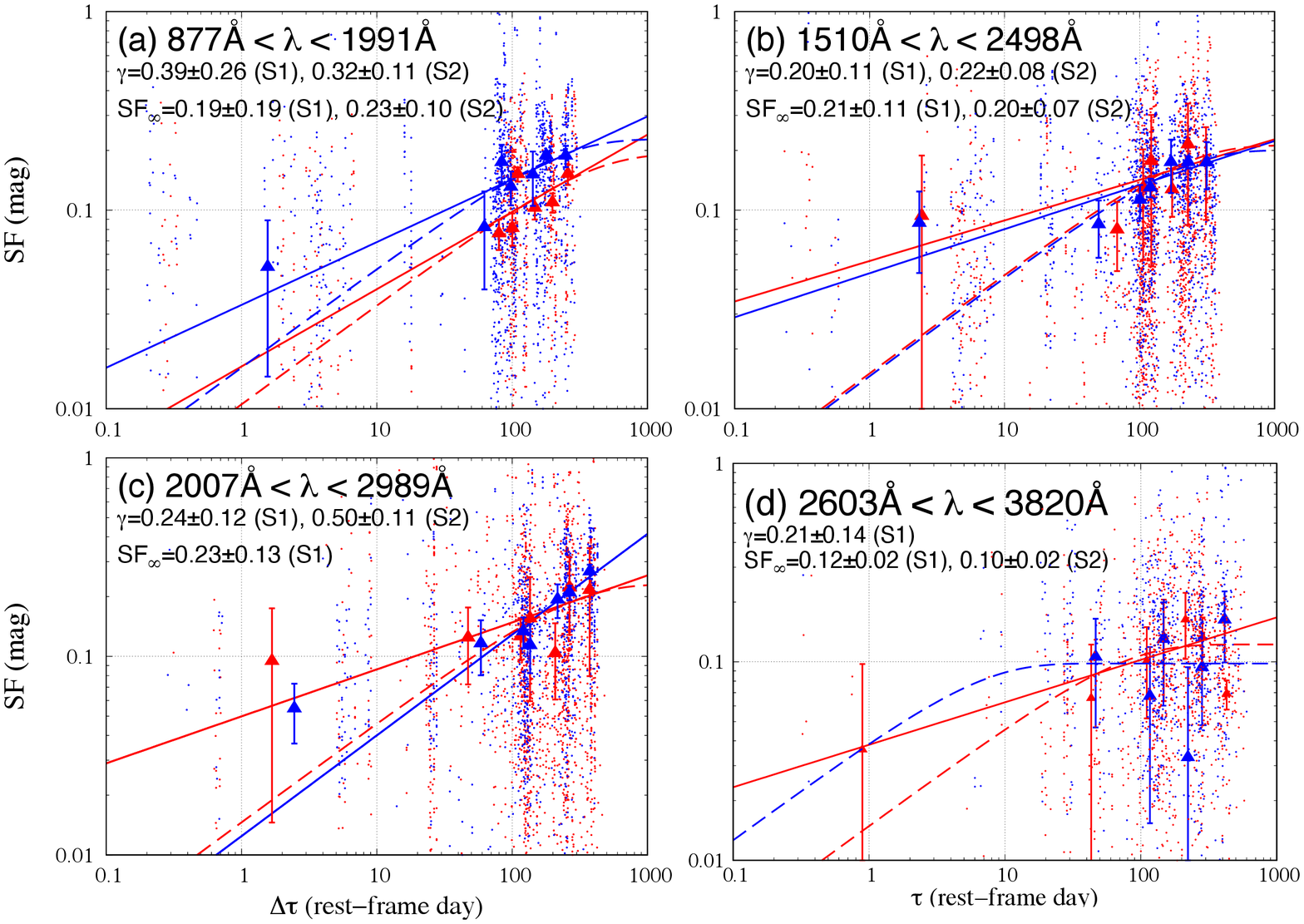}
\caption{Structure Functions for photometric data of the class S1 (red triangle) and S2 (blue triangle) 
quasars, by dividing the rest-frame wavelength into 4 ranges; (a) $877~{\rm \AA} < \lambda < 1991~ 
{\rm \AA}$, (b) $1510~{\rm \AA} < \lambda < 2498~{\rm \AA}$, (c) $2007~{\rm \AA} < \lambda 
< 2989~{\rm \AA}$, and (d) $2603~{\rm \AA} < \lambda < 3820~{\rm \AA}$. Red and blue dots 
represent the flux variability of 25 sample for all combinations of the observation epochs, 
corresponding to the classes S1 and S2 quasars. The structure functions of the classes S1 
(red lines) and S2 (blue lines) quasars are fitted with a power low ($solid~line$) and a DRW 
($dashed~line$). On short timescale ($<~$10 days), the structure function of the class S1 (or S2) 
quasars in panels (a) (or (d)) cannot be obtained due to larger photometric error than their mean 
flux variability. Fitting parameters of eqns. (6) and (7) are described in each panel.}
\end{figure*}

\begin{figure*}
\figurenum{6}
\plotone{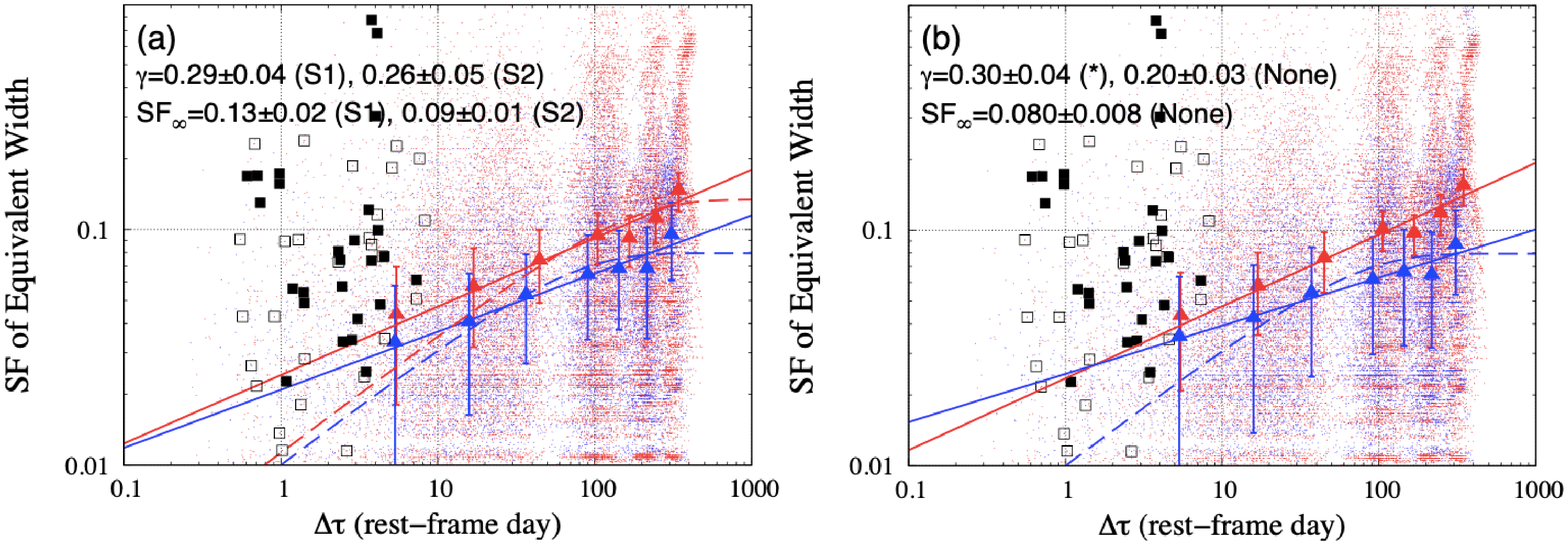}
\caption{The same as Figure 5, but for the structure functions of {C\,{\footnotesize IV}\ }BAL EWs; (a) for 
the classes S1 and S2 quasars, including both BAL[A] and BAL[B], (b) for the {C\,{\footnotesize IV}\ }BALs 
that presented short timescale variability marked with asterisks in Table 1 ($*$: red) or 
not (None: blue). Red and blue dots indicate the UDSFs of 25 sample with smaller uncertainties 
($\sigma_{\rm UDSF}<0.1$), corresponding to the classes S1 and S2 quasars. Filled (or open) black 
squares indicate short timescale significant variability of BAL[A] (or BAL[B]) defined by H19, 
and these plots are represented as UDSFs.}
\end{figure*}

\begin{figure}
\figurenum{7}
\plotone{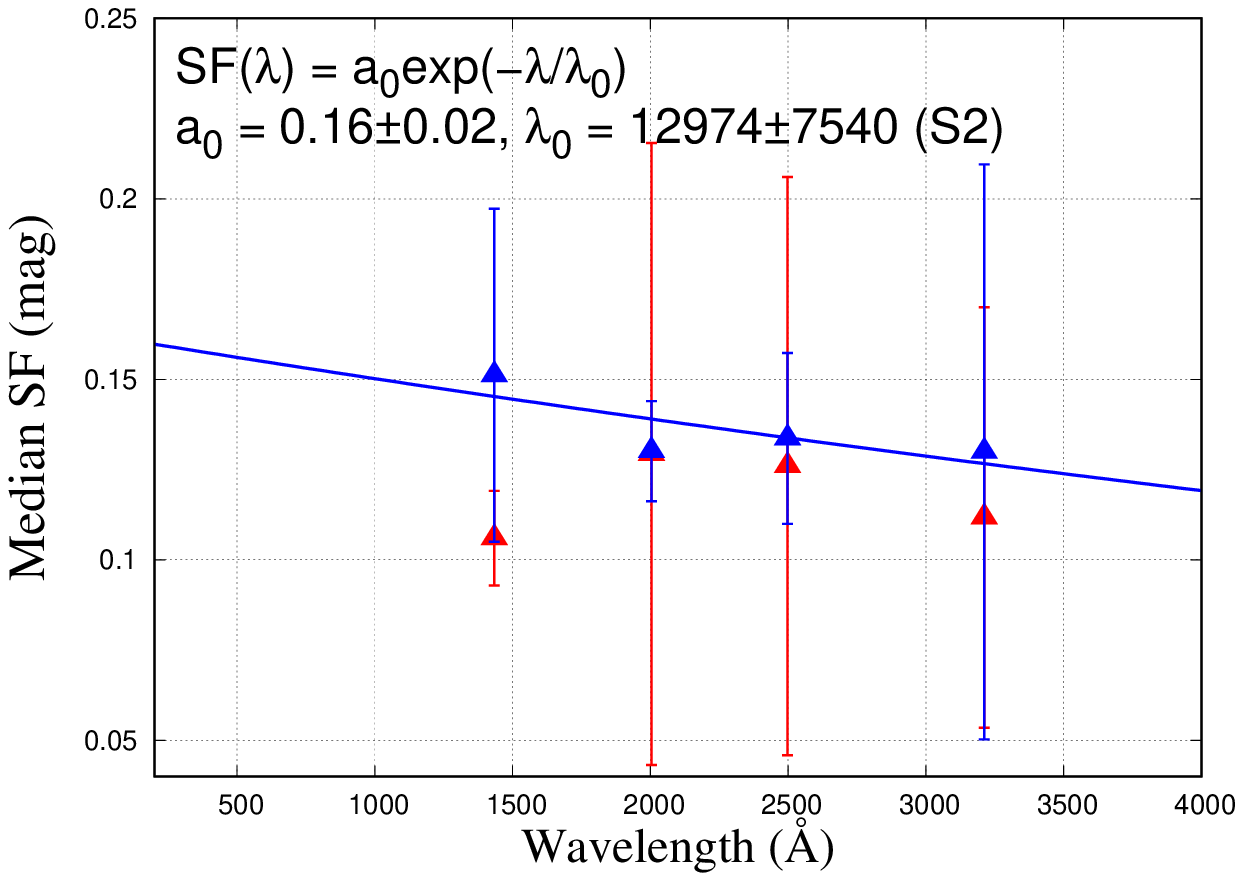}
\caption{The median structure functions as a function of the rest-frame wavelength 
for the class S1 (red triangle) and S2 (blue triangle) quasars. The structure
functions are the median values of each panel of Figure 5. The class S2 quasars are 
well fitted by eqn. (8). The fitting parameters are also shown.}
\end{figure}

\subsection{Correlation between {C\,{\footnotesize IV}\ }BAL EW variability and Physical Properties of 
Sample Quasars}

In order to probe the relation between BAL variability and physical properties 
of quasars, we also analysis the correlation between structure functions of 
{C\,{\footnotesize IV}\ }BAL EWs and physical parameters of the quasars such 
as bolometric luminosities $L_{\rm bol}$, black hole masses $M_{\rm BH}$, and 
Eddington ratios $L_{\rm bol}/L_{\rm Edd}$ after dividing into the classes S1 and S2 quasars. 
As listed in Table 1, we obtain these physical parameters from \citet{2019ArXiv190403199G} 
for 17 (bolometric luminosities), 15 (black hole masses) and 15 (Eddington ratios) out 
of the 25 quasars. Since the BAL variability is supposed to be sensitive to the 
ionization flux, we also study the relation between BAL variability and a physical 
quantity proportional to accretion disk temperature at an inner radius 
$T_{\rm in}~(\propto L_{\nu}^{3/8} M_{\rm BH}^{-3/4} \propto L_{\rm bol}^{3/8} M_{\rm BH}^{-3/4};$ 
see Section 4.1.3 and 4.2). For the 25 quasars, we assume the typical SED described by the 
bolometric corrections for $L_{\rm bol}=\zeta \lambda L_{\lambda}$, where $L_{\lambda}$ is 
a monochromatic luminosity, and factors $\zeta$ are 4.2, 5.2, and 8.1 for 1450, 3000, and 
5100 ${\rm \AA}$, respectively \citep{2012MNRAS...422..478R}.

Figure 8 summarizes the relations between BAL variability in each timescale 
and physical parameters. In relation to Figure 8, Table 2 lists the correlation 
coefficients with their $p$-values for each rest-frame time separation range (unit of 
days): $\Delta \tau<$ 10, $10<\Delta \tau<$ 30, $30<\Delta \tau<$ 100, and $100<\Delta \tau$. 
Based on Table 2, we denote the details of the correlation between BAL variability 
and physical parameters in the following sections. 

\subsubsection{$SF_{\rm BAL}$ vs. $L_{\rm bol}$: Figure 8a}
The classes S1 and S2 quasars do not show possible correlations 
($0.3<|r|$ or $0.3<|r|<0.4$ with $p$-value$>$0.05) for any time separations.  

\subsubsection{$SF_{\rm BAL}$ vs. $M_{\rm BH}$: Figure 8b}
Throughout, the class S1 quasars do not indicate possible correlations. 
On the other hand, the class S2 quasars show a moderate positive 
($r=0.47$ with $p$-value $=$ 0.03) or a strong positive correlation 
($r=0.67$ with $p$-value $=$ 0.02) after $\Delta \tau > 30$ days. 

\subsubsection{$SF_{\rm BAL}$ vs. $L_{\rm bol}/L_{\rm Edd}$: Figure 8c}
As a whole, the class S2 quasars present strong negative correlations ($r<-0.6$ with 
$p$-value $<$ 0.03). In contrast, the class S1 quasars show no correlations in any cases. 

\subsubsection{$SF_{\rm BAL}$ vs. $L_{\rm bol}^{3/8} M_{\rm BH}^{-3/4}$: Figure 8d}
The class S2 quasars show strong negative correlations for $\Delta \tau >$ 10 days, while 
the class S1 quasars do not exhibit possible correlations. \\

There are no possible correlations between BAL variability and physical parameters for the 
class S1 quasars, and the overall distribution of the class S1 (14 quasars) and S2 (11 quasars) 
quasars ($|r_{\rm All}|<0.3$). Here, we emphasize that RM 613 and RM 770 whose Eddington 
ratios are comparably higher in the class S1 quasars ($L_{\rm bol}/L_{\rm Edd}=0.245\pm0.005$, 
and $0.344\pm0.007$, respectively) exhibit a different trend from the class S2 quasars with 
the strong negative correlation between {C\,{\footnotesize IV}\ }BAL variability and Eddington 
ratios (for $L_{\rm bol}/L_{\rm Edd} <$ 0.2).

\begin{figure*}
\figurenum{8}
\plotone{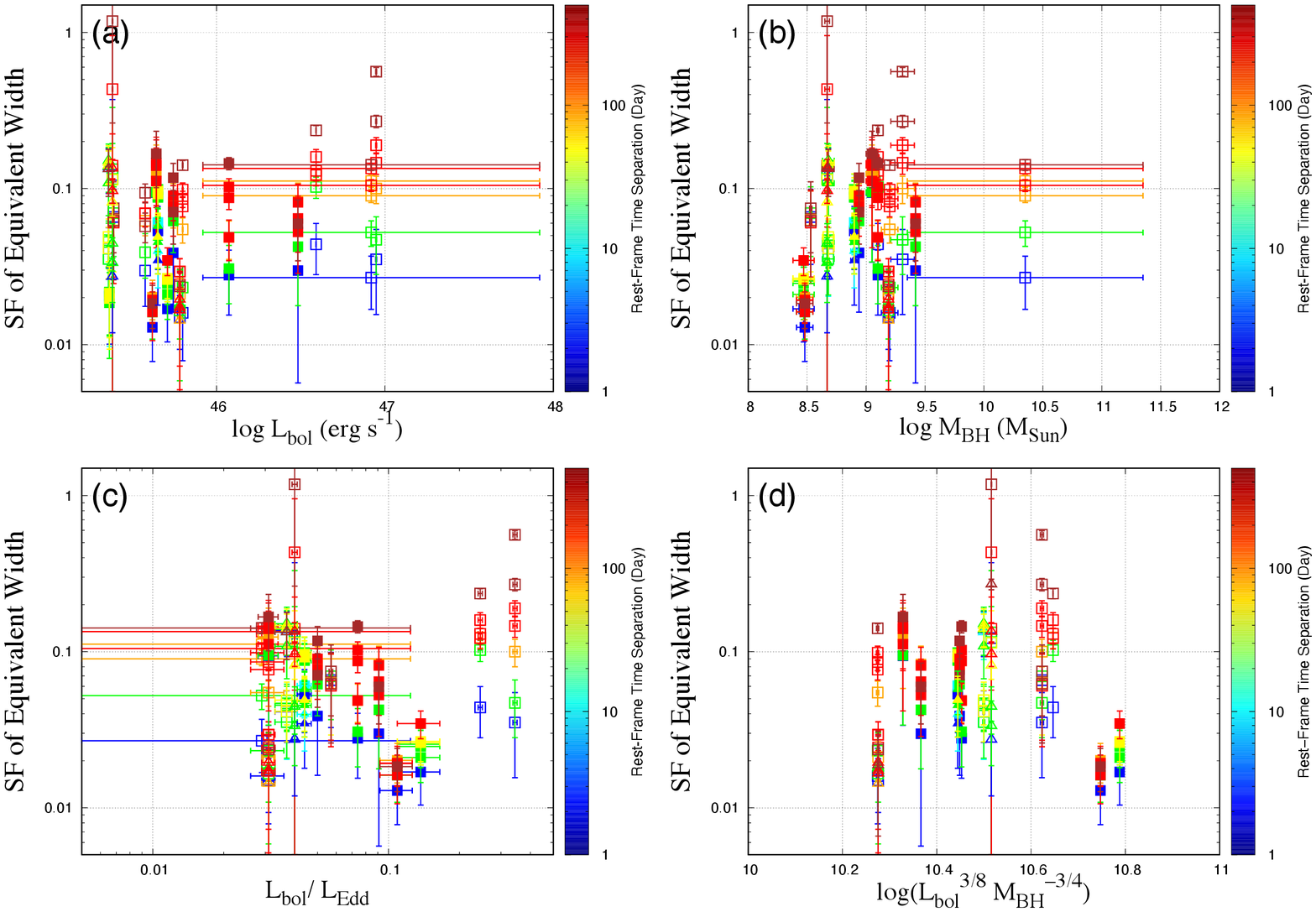}
\caption{The relation between structure functions of {C\,{\footnotesize IV}\ }BAL EWs 
and physical properties at each rest-frame time-lag. Each panels indicate structure 
functions of BAL[A][B] for individual quasars based on eqn. (4) versus (a) bolometric luminosity, 
(b) black hole mass, (c) Eddington ratio, and (d) temperature of the accretion disk based on the 
standard model. Open and filled squares (or triangles) show the distribution for BAL[A] 
(or BAL[B]) of the classes S1 and S2 quasars. Colors indicate rest-frame time separation, 
where redder colors correspond to larger time lag, as shown in the color keys. 
}
\end{figure*}

\begin{deluxetable*}{ccccc}[b!]
\tablecaption{The correlation coefficient with $p$-values between BAL variability and physical parameters \label{tab:mathmode}}
\tablecolumns{4}
\tablenum{2}
\tablewidth{0pt}
\tablehead{
\colhead{The range of time separation $\Delta \tau$ (days)}&
\colhead{$r_{\rm S1}$$^a$ ($p$-values): 14 quasars}&
\colhead{$r_{\rm S2}$$^b$ ($p$-values): 11 quasars}&
\colhead{$r_{\rm All}$$^c$ ($p$-values): 25 quasars}&
}
\startdata
\multicolumn{4}{c}{$SF_{\rm BAL}$ vs. $L_{\rm bol}$ (Figure 8a)} \\ \hline 
$\Delta \tau<$ 10 & -0.373 (0.231) & -0.083 (0.831) & -0.297 (0.190) \\ 
$10<\Delta \tau<$ 30& -0.202 (0.392) & 0.135 (0.569) & -0.122 (0.452)\\ 
$30<\Delta \tau<$ 100& 0.203 (0.467) &  0.322 (0.283) & 0.255 (0.189)\\ 
$100<\Delta \tau$& 0.060 (0.722) & -0.051 (0.825) & 0.062 (0.646) \\ 
For all data of $\Delta \tau$ & 0.064 (0.560) & 0.163 (0.203) & 0.088 (0.291) \\ \hline
\multicolumn{4}{c}{$SF_{\rm BAL}$ vs. $M_{\rm BH}$ (Figure 8b)} \\ \hline 
$\Delta \tau<$ 10 & -0.481 (0.134) & 0.382 (0.349) & -0.268 (0.266) \\ 
$10<\Delta \tau<$ 30 & -0.320 (0.182) & 0.461 (0.072) & -0.118 (0.497) \\ 
$30<\Delta \tau<$ 100& 0.039 (0.894) & \bf{0.671 (0.023)} & 0.228 (0.271) \\ 
$100<\Delta \tau$& -0.142 (0.429) & \bf{0.475 (0.029)} & -0.044 (0.753) \\
For all data of $\Delta \tau$ & -0.080 (0.489) & \bf{0.495 (0)} & 0.010 (0.909) \\ \hline
\multicolumn{4}{c}{$SF_{\rm BAL}$ vs. $L_{\rm bol}/L_{\rm Edd}$ (Figure 8c)} \\ \hline 
$\Delta \tau<$ 10 & -0.098 (0.774) & \bf{-0.765 (0.026)} & -0.167 (0.493) \\ 
$10<\Delta \tau<$ 30& 0.005 (0.984) & \bf{-0.754 (0)} & -0.144 (0.407)\\ 
$30<\Delta \tau<$ 100& 0.267 (0.355) & \bf{-0.737 (0.009)} & 0.062 (0.769)\\ 
$100<\Delta \tau$& 0.208 (0.244) & \bf{-0.825 (0)} & 0.197 (0.153)\\ 
For all data of $\Delta \tau$ & 0.187 (0.103) & -\bf{0.670 (0)} & 0.146 (0.093) \\ \hline
\multicolumn{4}{c}{$SF_{\rm BAL}$ vs. $L_{\rm bol}^{3/8} M_{\rm BH}^{-3/4}$ (Figure 8d)} \\ \hline 
$\Delta \tau<$ 10 & 0.377 (0.253) & -0.687 (0.059) & 0.093 (0.705)\\ 
$10<\Delta \tau<$ 30& 0.326 (0.173) & \bf{-0.687 (0.003)} & -0.025 (0.886)\\ 
$30<\Delta \tau<$ 100& 0.112 (0.703) & \bf{-0.817 (0.002)} & -0.207 (0.320)\\ 
$100<\Delta \tau$& 0.241 (0.177) & \bf{-0.751 (0)} & 0.107 (0.443)\\ 
For all data of $\Delta \tau$ & 0.181 (0.115) & -\bf{0.688 (0)} & 0.036 (0.677)\\
\enddata
\tablenotetext{a}{The correlation coefficient for the class S1 quasars.}
\tablenotetext{b}{The correlation coefficient for the class S2 quasars.}
\tablenotetext{c}{The correlation coefficient for the overall distribution of the classes S1 and S2 quasars.}
\tablecomments{The correlations with $|r|>0.3$ and $p$-value$<$0.05 are marked in bold face. The 
$p$-values with zero values indicate $p$-value$<$0.0001.}
\end{deluxetable*}

\section{Discussion} 
\subsection{Ionization state change in Outflows}
We first consider whether a change in the ionization condition of outflow winds 
explains the observational results. 

\subsubsection{Correlation between the BAL EWs and quasar variability}

\citet{2013A&A...557A..91T} found a clear correlation between variability of 
{C\,{\footnotesize IV}\ }BAL EWs and the rest-frame UV flux variability in a gravitational 
lensed BAL quasar APM 08279+5255, supporting the ionization state change as a 
possible origin of the BAL variability. \citet{2019MNRAS.468..2379V} also found 78 
quasars of the SDSS DR12 with high signal-to-noise ratio spectra show a weak/moderate 
correlation between BAL and flux variability taken by the SDSS and the Catalina 
Real-Time Transient Survey (CRTS) catalog. \citet{2017ApJS...229..22H} analyzed the 
multi-epoch SDSS DR12 spectra of BAL quasars, and using EW ratios ($R$) of 
{Si\,{\footnotesize IV}\ }to {C\,{\footnotesize IV}\ }BAL, they suggested that a recombination 
timescale of BAL clouds is only a few days. This means the typical absorber density 
$n_e$ is $\sim 10^6$ cm$^{-3}$. \citet{2019arXiv190803844L} discovered a clear 
correlation between BAL variability and flux changes for 21 BAL quasars from the 
SDSS-I/II/III, including some of our sample (RM 357, RM 722, RM 729, RM 730, and 
RM 786). These studies conclude that changes in ionizing continuum 
cause BAL variability in a various time scale.

Most of distributions in Figure 3 show no positive or negative correlations. Meanwhile, 
BAL EW variability in Figure 4 shows weak, moderate positive ($r=0.39$ and 
$r_{\Delta \tau < 5}=0.42$  for RM 284 [A], and $r=0.34$ for RM 717 [A]) or a strong 
positive ($r > 0.7$ for RM 730 [B]) correlation with $R$-band variability in the timescale 
from $<$10 days to longer; these {C\,{\footnotesize IV}\ }BAL EWs increase (or decrease) 
when these quasars dim (or brighten) in short timescales.  
This result suggests a recombination time of outflow gas clouds is within a few days or a week. 
In contrast to the preceding argument, 
RM 339 [A], RM 730 [A], RM 770[A] do not indicate short timescale BAL variability. 
Moreover, relatively bright quasars RM 730 ($m_i=17.98$) and RM 770 ($m_i=16.46$ and 
log$L_{bol}=46.948\pm0.003$) show no significant $R$-band variability (Figure 1). This trend 
likely follows a negative correlation between variability amplitude and the quasar luminosity 
\citep[e.g,][]{1999MNRAS.306..637G,2004ApJ...601..692V,2012ApJ...753..106M}. Thus, 
we speculate BAL variability is mainly caused by quasar flux variability\footnote{BALs could be 
changed even if the ionization flux is constant if they originate in the fast unsteady flows 
\citep{2012ASPC..460..171P}.}. Especially, less luminous quasars tend to show the correlation 
between BAL and flux variability. In terms of RM 339 [A], RM 730 [A], and RM 770 [A], 
it is necessary to consider another mechanism to change BAL EWs (see Section 4.2). 

\subsubsection{Structure Function}
The structure functions in Figure 5 show following trends; (i) variability 
amplitudes become slightly larger with time separations as suggested in previous studies 
\citep[e.g.,][]{2004ApJ...601..692V,2010ApJ...721..1024M}, and power-law indexes $\gamma$ in 
each panel are consistent with those of \citet{2004ApJ...601..692V}, 
(ii) DRW model is not well fitted at short-timescale, as already reported by 
\citet{2012ApJ...753..106M}. In other words, the classes S1 and S2 quasars in this study
have consistent properties of flux variability with samples in previous studies.

Based on the ionization state change scenario in outflows, the larger flux changes in quasar 
yield, the larger EW variability in BALs. According 
to the scenario, the class S1 quasars, with larger BAL variability 
amplitudes than the class S2 quasars (Figures 6a and 6b), would show relatively larger 
flux variability than the class S2 quasars. However, contrary to this expectation, Figure 5 
indicates almost comparable flux variability amplitudes with the classes S1 and S2 
quasars in each panel. 

In Figures 6a and 6b, we conjecture another possibility for the systematic difference in the 
BAL variability trend between the class S1 and S2 quasars (or the categories “$*$'' and “None''); 
the BAL variability follows the trend that absorption lines with shallower line profiles (smaller EWs) 
more easily show the larger fractional variability $|\Delta EW/\bigl<EW\bigr>|$ 
\citep[e.g,][]{2007ApJ...656..73L}. However, there is no significant difference in $\bigl<EW\bigr> $ 
between the class S1 (15.1~$\pm$~0.4~${\rm \AA}$) and S2 quasars (16.2~$\pm$~0.3~${\rm \AA}$) 
\footnote{Average EW for the classes S1 and S2 quasars based on Table 3 in H19.}. 
Namely, the difference of their average EW ($\bigl<EW\bigr>$), which could cause BAL 
variability easily due to the ionization state change, can not be the origin behind the difference between 
the classes S1 and S2 quasars. Furthermore, in Figures 6a and 6b, significant BAL variability 
($<$10-day) for BAL[A] show weak negative correlation between rest-frame time lag and 
$UDSF(< 0.2)$ with a correlation coefficient of $r=-0.38$. The trend also suggests that BAL 
variability requires another physical process other than flux variability (see also Section 4.2). 

\subsubsection{Relation between BAL Variability and Physical Properties}
In general, flux variability amplitude decreases with quasar luminosities, 
increases with black hole masses and decreases with Eddington ratios as already suggested 
in previous studies \citep[e.g.,][]{2007MNRAS.375..989W,2008MNRAS.383.1232W,
2013A&A...560A.104M,2018ApJ...854..160R} as explained in Section 4.1.1. If BAL 
variability is mainly driven by flux changes, the variability timescales and/or amplitude would 
also depend on physical properties of quasars. 

We also evaluate the relation between BAL variability and the accretion 
disk temperature at an inner radius $T_{\rm in}$. Based on the standard thin accretion disk model  
\citep{1973A&A...24...337S}, the effective temperature $T_{\rm eff}$ is proportional 
to $(M_{\rm BH} \dot{M}/R^3)^{1/4}$, where $\dot{M}$ and $R$ are mass accretion rate and 
disk radius, respectively. For example, at the region of inner radius $R=3R_{\rm S}$ (Schwarzschild 
radius $R_{\rm S}=2GM_{\rm BH}/c^2 \propto M_{\rm BH}$), the disk temperature $T_{\rm in}$ is 
proportional to $(\dot{M}/M_{\rm BH}^2)^{1/4}$. The disk luminosity at optical/UV $L (=\nu L_{\nu})$, 
$\dot{M}$ and $M_{\rm BH}$ are related with $L \propto (\dot{M} M_{\rm BH})^{2/3}$ 
\citep[i.e., $\dot{M} \propto L^{3/2}/M_{\rm BH}$; ][]{2004A&A...426..797}. 
Therefore, the disk temperature $T_{\rm in}$ is described by 

\begin{equation}
T_{\rm in} \propto L^{3/8} M_{\rm BH}^{-3/4} \propto L_{\rm bol}^{3/8} M_{\rm BH}^{-3/4}.
\end{equation}
The negative correlation between BAL variability and the disk temperature $T_{\rm in}$ would 
also be expected, since, as well as the Eddington ratio ($\propto L_{\rm bol} M_{\rm BH}^{-1}$), 
the $T_{\rm in}$ is associated with bolometric luminosity and black hole mass. In fact, the class S2 
quasars only show the strong correlation between structure functions of {C\,{\footnotesize IV}\ }BAL 
and physical properties of the Eddington ratio and $L_{\rm bol}^{3/8} M_{\rm BH}^{-3/4}$ (negative 
correlation) in Figure 8. These BAL variability trends are consistent with the relation between flux 
variability and physical parameters of quasars. The results also support the ionization state 
change scenario and are contrary to the cloud crossing scenario. 

As described in Section 3.3, two quasars at higher Eddington ratios (the class S1 quasars 
RM 613 and RM 770) exhibit relatively large {C\,{\footnotesize IV}\ }BAL variability. 
The average {C\,{\footnotesize IV}\ }BAL EWs of these two quasars are smaller 
(4.8~$\pm$~0.2~${\rm \AA}$ for RM 613, and 3.6~$\pm$~0.1~${\rm \AA}$  
for RM 770) than those of the other quasars (15.1~$\pm$~0.4~${\rm \AA}$ as the average 
EW of the class S1 quasars). The {C\,{\footnotesize IV}\ }BAL variability trend of RM 613 
and RM 770 suggests the observational trend that shallower intrinsic absorption lines more 
easily show larger BAL variability as explained in Section 4.1.2. On the other hand, at lower 
Eddington ratios ($L_{\rm bol}/L_{\rm Edd} < 0.1$), the class S1 quasar RM 217 ($\bigl<EW\bigr>
=$ 4.2~$\pm$~0.4~${\rm \AA}$ for BAL[A]) presents the largest BAL variability, while 
RM 257 [A] with a relatively large average EW ($\bigl<EW\bigr>=$ 42.6~$\pm$~0.4~
${\rm \AA}$) exhibits the smallest BAL variability in the class S1 quasars. 
As a whole, the class S1 quasars have a larger sample variance of the average 
{C\,{\footnotesize IV}\ }BAL EW than that of the class S2 quasars except for the data of 
RM 039 with $\bigl<EW\bigr>$ of 88.5 $\pm$ 0.7~${\rm \AA}$ (see APPENDIX A and Figure A 
in this paper). Consequently, the variance of the average EW in the class S1 quasars yields 
no correlations in Figure 8\footnote{Namely, larger variance of $\bigl<EW\bigr>$ is not 
suitable for statistical analysis.}. 

In comparable EW of BAL samples, the correlation between BAL variability and  
black hole masses (positive correlation), and Eddington ratios or accretion disk 
temperature (negative correlation) indicates that the BAL variability amplitude depends 
on the properties of the accretion disk. Quasars with lower Eddington ratios frequently 
change the accretion rates (i.e., unstable fuel supply onto accretion disk). The frequent change 
of accretion rate causes rapid and large flux variability of quasars 
\citep[c.f.,][]{2008MNRAS.383.1232W,2018ApJ...854..160R}. Consequently, the ionization states 
in outflow winds can be affected by flux variability in quasars with lower Eddington ratios 
because of a sudden increasing or decreasing of a fuel supply onto the accretion disk. 

\subsection{The Changes in Shielding Gas?} 
We find several important results; (1) Among the $five~quasars$, three quasars present weak, 
moderate, or a strong positive correlation between the iPTF $R$-band variability and BAL 
variability (Figure 4), (2) there is almost no significant difference of flux variability amplitudes 
between BAL quasars with the significant EW variability and the others (Figure 5), while BAL 
variability for the class S1 are systematically larger than that for the class S2 from $<10$~days 
to longer timescales ($>10$~days) (Figure 6), and (3) BAL variability and physical parameters 
show a possible positive (black hole masses) and negative correlations (Eddington ratios and 
the accretion disk temperature) in the class S2 quasars (Figure 8). From Figure 6, the classes S1 
and S2 quasars could show the systematic difference of flux variability between them based on 
the ionization state change. However, the result (2) does not show such a trend, and 
this fact indicates that the BAL variability cannot be explained only by the flux variability in quasars. 
 
As an ancillary mechanism to explain short and/or long timescale BAL variability, 
the variations in shielding gas is proposed \citep[c.f.,][]{2013MNRAS.435..133H}; it is a inner 
disk flow that fails to escape from the disk (i.e., failed wind) and presents frequent variability in very 
short timescales of the order of days to weeks 
\citep{2000ApJ...543..686P,2004ApJ...616..688P,2012ASPC..460..171P,2019AA...630..A94G}. The 
shielding gas appears to prevent over-ionization of outflows at a down stream. If the shielding gas 
locating at the innermost of an accretion disk fluctuates its own ionization state, an amount of ionizing 
photons (e.g, EUV continuum) is probably adjusted due to variations of the shielding gas. Consequently, 
outflows change their ionization state. As an observational suggestion, broad absorbing features via 
X-ray ultra-fast outflows have rapidly changed in a timescale of hours to months that is consistent with 
a typical timescale of the shielding gas 
\citep[][see also Giustini \& Proga 2019]{2009ApJ...697...194S,2014ApJ...784..77G}. 

One of the candidate of shielding gas is a warm absorber that has been detected as 
absorption edges in X-ray spectra \citep[e.g.,][]{2002ApJ...567...37G, 2006ApJ...644..709G}. 
Warm absorbers were originally proposed to avoid over-ionization of the 
outflows \citep{1995ApJ...451..498}. In most case, strong X-ray warm absorbers are detected in 
spectra of BAL quasars. To put it another way, strong X-ray absorption is rarely detected in 
spectra of non-BAL quasars \citep[e.g., HS 1700+6416;][]{2012ApJ...544..A2L}.
Notably, the result (2) indicates that the effect of variations in shielding gas is probably more 
prominent in the class S1 quasars, since the BAL variability timescales of this class 
are similar to that of the shielding gas \citep{2012ASPC..460..171P}.   
If we detect simultaneous changes of X-ray absorbers and BAL variability, the ionization state 
change scenario becomes more robust the physical 
mechanism of BAL variability. 

\section{Conclusion} 

We examined the correlation between the variability of flux for BAL quasars and their 
EWs in a various time scales from $<10$~days to a few years in the quasar rest-frame. 
In the study of the correlation, we used the data sets of EW of BALs taken by the Sloan 
Digital Sky Survey Reverberation Mapping (SDSS-RM) project, reported by H19 and 
photometric data taken by the iPTF) with $g$, $R$-band and the PS1 with $grizy$ bands. 
We divide the sample into two the classes (S1 and S2) according to the presence or the lack of 
short timescale BAL variability. Our results are summarized as follows. 
 \begin{itemize}
      \item[(1)] Among the $five~quasars$ that satisfy the conditions (1) and (2) in Section 2.2.1, 
                     three quasars present weak, moderate (for RM 284 and RM 717), and a strong 
                     correlation (for RM 730) between flux variability and BAL variability (Figure 4).
      \item[(2)] The classes S1 and S2 quasars show no significant difference in flux variability amplitudes (Figure 5). 
      \item[(3)] There is systematic difference in BAL variability amplitudes between the classes S1 
                     and S2 quasars (Figure 6). 
      \item[(4)] The distributions of BAL variability amplitudes and black hole masses, Eddington ratios and the
                      accretion disk temperature exhibit strong negative correlations for the class S2 quasars (Figure 8). 
      \end{itemize}
These results indicate that BAL variability primarily requires variations in the ionizing continuum, 
and secondarily an ancillary mechanism such as variability in shielding gas except for the gas motion 
scenarios.

\acknowledgments
We are very grateful to the staff in Ishigakijima Astronomical Observatory and Yaeyama Star Club 
for supporting to prepare the environment to describe this paper. We would like to acknowledge the 
anonymous referee for useful comments. 

Funding for the Sloan Digital Sky Survey IV has been provided by the Alfred P. Sloan Foundation, 
the U.S. Department of Energy Office of Science, and the Participating Institutions. SDSS acknowledges 
support and resources from the Center for High-Performance Computing at the University of Utah. 
The SDSS web site is www.sdss.org. SDSS is managed by the Astrophysical Research Consortium for 
the Participating Institutions of the SDSS Collaboration including the Brazilian Participation Group, the 
Carnegie Institution for Science, Carnegie Mellon University, the Chilean Participation Group, the French 
Participation Group, Harvard-Smithsonian Center for Astrophysics, Instituto de Astrofísica de Canarias, 
The Johns Hopkins University, Kavli Institute for the Physics and Mathematics of the Universe (IPMU) / 
University of Tokyo, the Korean Participation Group, Lawrence Berkeley National Laboratory, Leibniz 
Institut für Astrophysik Potsdam (AIP), Max-Planck-Institut für Astronomie (MPIA Heidelberg), Max-Planck-Institut für 
Astrophysik (MPA Garching), Max-Planck-Institut für Extraterrestrische Physik (MPE), National Astronomical 
Observatories of China, New Mexico State University, New York University, University of Notre Dame, 
Observatório Nacional / MCTI, The Ohio State University, Pennsylvania State University, Shanghai 
Astronomical Observatory, United Kingdom Participation Group, Universidad Nacional Autónoma de 
México, University of Arizona, University of Colorado Boulder, University of Oxford, University of Portsmouth, 
University of Utah, University of Virginia, University of Washington, University of Wisconsin, Vanderbilt 
University, and Yale University. 

This study has made use of the NASA/IPAC Infrared Science Archive, which is operated by the Jet Propulsion 
Laboratory, California Institute of Technology, under contract with the National Aeronautics and Space 
Administration. 

The intermediate Palomar Transient Factory (iPTF) project is a scientific collaboration among the California 
Institute of Technology, Los Alamos National Laboratory, the University of Wisconsin (Milwaukee), the 
Oskar Klein Center, the Weizmann Institute of Science, the TANGO Program of the University System 
of Taiwan, and the Kavli Institute for the Physics and Mathematics of the Universe. LANL participation 
in iPTF was funded by the U.S. Department of Energy as part of the Laboratory Directed Research and 
Development program. Part of this research was carried out at the Jet Propulsion Laboratory, California 
Institute of Technology, under a contract with NASA.

The Pan-STARRS1 Surveys (PS1) and the PS1 public science archive have been made 
possible through contributions by the Institute for Astronomy, the University of Hawaii, the Pan-STARRS 
Project Office, the Max-Planck Society and its participating institutes, the Max Planck Institute for Astronomy, 
Heidelberg and the Max Planck Institute for Extraterrestrial Physics, Garching, The Johns Hopkins 
University, Durham University, the University of Edinburgh, the Queen's University Belfast, the 
Harvard-Smithsonian Center for Astrophysics, the Las Cumbres Observatory Global Telescope 
Network Incorporated, the National Central University of Taiwan, the Space Telescope Science Institute, 
the National Aeronautics and Space Administration under Grant No. NNX08AR22G issued through the 
Planetary Science Division of the NASA Science Mission Directorate, the National Science Foundation 
Grant No. AST-1238877, the University of Maryland, Eotvos Lorand University (ELTE), the Los Alamos 
National Laboratory, and the Gordon and Betty Moore Foundation.

\appendix

\section{Relation between BAL Variability, Eddington Ratio, and Average EW}
We investigate the relation between BAL variability and Eddington ratio, and 
the average EW with following distributions:  
(i) Eddington ratio versus average EW, (ii) BAL variability versus Eddington ratio, and 
(iii) BAL variability versus average EW in Figure A. BALs of the class S1 
quasars with comparably smaller average EW, RM 217, RM 613, and RM 770 
(whose average EW are $\bigl<EW\bigr> = 4.2 \pm 0.4~{\rm \AA}, 4.8 \pm 0.2~{\rm \AA}$, 
and $3.6 \pm 0.1~{\rm \AA}$, respectively) exhibit extremely large BAL variability in the distributions 
(ii) and (iii). In contrast, those for RM 257 [A] with $\bigl<EW\bigr>=~ $42.6~$\pm$~0.4$~{\rm \AA}$ 
show the smallest BAL variability in the class S1 quasars. The sample variance of $\bigl<EW\bigr>$ 
for the class S1 quasar ($s^2 = 158.4$ ${\rm \AA}^2$) is larger than that for the class S2 quasars 
($s^2 = 10.5$ ${\rm \AA}^2$) except for RM 039 with $\bigl<EW\bigr>$ 
of 88.5 $\pm$ 0.7~${\rm \AA}$. As a result, the class S1 quasars seem to have no correlation 
($r_{\rm S1} =$ 0.19 and -0.29 for the distributions (ii) and (iii)). In terms of the class S2 quasars, 
they indicate negative correlations in the distributions (ii) ($r_{\rm S2} = -0.67$) and (iii) 
($r_{\rm S2} = -0.46$). Especially in the distribution (ii), despite comparable average EWs for RM 
116 [A], RM 339 [A], RM 408 [A], and RM 729 [A] ($\bigl<EW\bigr> = 8.2 \pm 0.3~{\rm \AA}, 
9.7 \pm 0.3~{\rm \AA}, 9.8 \pm 0.2~{\rm \AA}$, and $11.8 \pm 0.3~{\rm \AA}$, respectively), 
they suggest a negative correlation between BAL variability and Eddington ratios. This result 
indicates that the Eddington ratio can be an index of 
BAL variability. The negative correlation between the average EWs and BAL 
variability (distribution (iii); $r=-0.46$ for the class S2 quasars) means the trend that shallower 
absorption lines show larger fractional variability \citep[e.g,][]{2007ApJ...656..73L}.

\begin{figure}[b]
\figurenum{A}
 \plotone{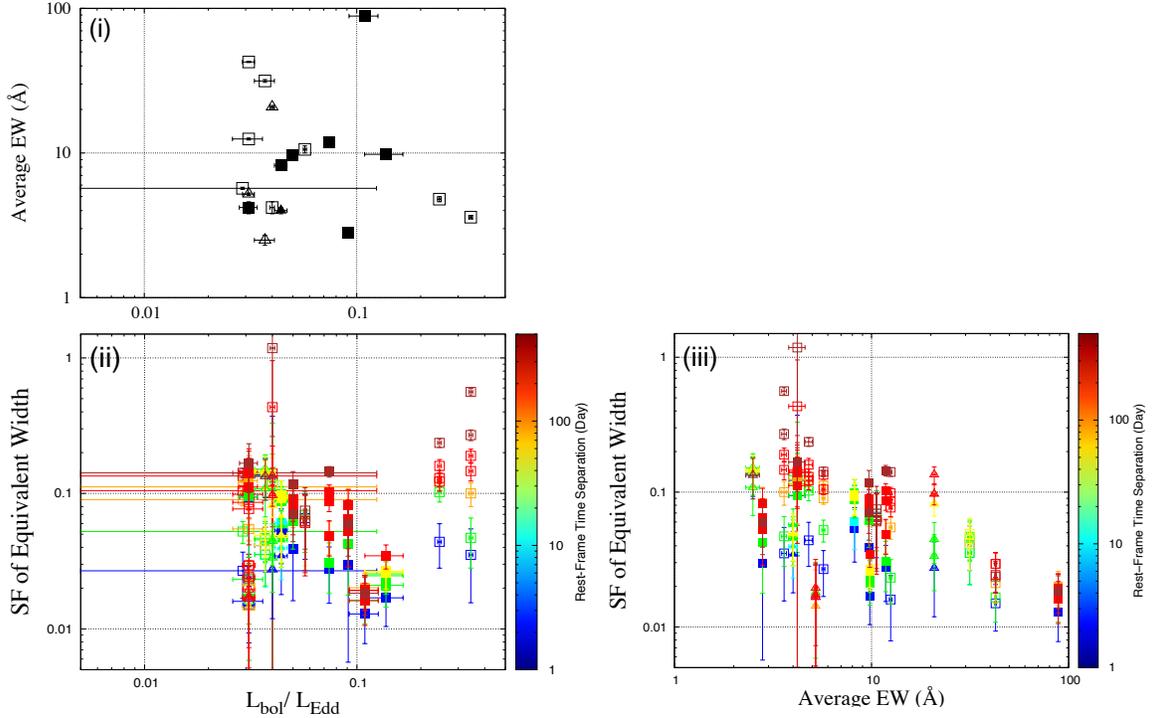}
\caption{
The plot of Eddington ratios versus average EWs ($top$), 
Eddington ratios versus BAL variability (the same as Figure 8c; $bottom~left$), and 
average EWs versus BAL variability ($bottom~right$). 
The symbols (Open and filled squares or triangles) are the same in Figure 8. 
Colors for bottom panels indicate the rest-frame time separation, where redder colors 
correspond to larger time lag, as shown in the color keys.
}
\end{figure}



\end{document}